%% file: mu-e-nnlo.tex
\documentclass[11pt,a4paper]{article}
\usepackage[left=2.35cm,top=3cm,bottom=3cm,right=2.3cm]{geometry}
\usepackage{amsmath}
\usepackage{amssymb}
\usepackage{upgreek}
\usepackage{cite}
\usepackage{graphicx}
\usepackage[caption=false]{subfig}
\usepackage{multirow}
\usepackage{mcmule}
\usepackage[colorlinks=true
,urlcolor=blue
,anchorcolor=blue
,citecolor=blue
,filecolor=blue
,linkcolor=blue
,menucolor=blue
,linktocpage=true
,pdfproducer=medialab
]{hyperref}
\usepackage{tikz}
\usetikzlibrary{decorations.pathmorphing,positioning,decorations.pathreplacing,shapes,calc}
\usetikzlibrary{decorations.pathreplacing}
\usetikzlibrary{decorations.markings}
\usetikzlibrary{arrows}

\renewcommand{\Re}{\mathrm{Re}}

\newcommand{\cM}{\mathcal{M}}

\def\MeV{{\rm MeV}}

\newcommand{\pp}{\phantom{(0)}}
\newcommand{\po}{\phantom{0}}

\newcommand{\Amp}[2]{\mathcal{A}_{#1}^{(#2)}}

\newcommand{\szt}{\sigma_0}
\newcommand{\soe}{\sigma^{(1)}_e}
\newcommand{\sox}{\sigma^{(1)}_{e\mu}\,\Bigl\{^-_+}
\newcommand{\som}{\sigma^{(1)}_\mu}
\newcommand{\sol}{\sigma^{(1)}_{\rm lep}}
\newcommand{\soh}{\sigma^{(1)}_{\rm had}}
\newcommand{\sot}{\sigma_1\,\Bigl\{^-_+}
\newcommand{\ste}{\sigma^{(2)}_e}
\newcommand{\stx}{\sigma^{(2)}_{e\mu}\,\Bigl\{^-_+}
\newcommand{\stm}{\sigma^{(2)}_\mu}
\newcommand{\stl}{\sigma^{(2)}_{\rm lep}\,\Bigl\{^-_+}
\newcommand{\sth}{\sigma^{(2)}_{\rm had}\,\Bigl\{^-_+}
\newcommand{\stt}{\sigma_2\,\Bigl\{^-_+}

\begin{document}
\thispagestyle{empty}

\begin{flushright} \small
FR-PHENO-2022-11;
IPPP/22/81;
MPP-2022-139;
PSI-PR-22-36;
ZU-TH 59/22;
UWThPh-2022-17
\end{flushright}
\vspace{1em}
\begin{center}
    {\Large\bf Muon-electron scattering at NNLO}
\\
\vspace{1em}
{\sc
    A.\,Broggio$^{a}$,
    T.\,Engel$^{b,c,d}$,
    A.\,Ferroglia$^{e,f}$,
    M.\,K.\,Mandal$^{g,h}$,
    P.\,Mastrolia$^{i,g}$,
    M.\,Rocco$^{b}$,
    J.\,Ronca$^{j}$,
    A.\,Signer$^{b,c}$,
    W.\,J.\,Torres Bobadilla$^{k}$,
    Y.\,Ulrich$^{l}$,
    M.\,Zoller$^{b}$
}\\[2em]
{\sl
${}^a$ Faculty of Physics, University of Vienna, Boltzmanngasse 5, A-1090 Vienna, Austria\\
\vspace{0.3cm}
${}^b$ Paul Scherrer Institut,
CH-5232 Villigen PSI, Switzerland \\
\vspace{0.3cm}
${}^c$ Physik-Institut, Universit\"at Z\"urich,
Winterthurerstrasse 190,
CH-8057 Z\"urich, Switzerland\\
\vspace{0.3cm} ${}^d$ Albert-Ludwigs-Universit\"at Freiburg,
Physikalisches Institut, \\ Hermann-Herder-Stra{\ss}e 3, D-79104
Freiburg, Germany \\
\vspace{0.3cm} ${}^e$ Physics Department, New York City College of
Technology, The City University of New York,\\ 300 Jay Street,
Brooklyn, NY 11201, USA\\
\vspace{0.3cm} ${}^f$ The Graduate School and University Center, The
City University of New York,\\ 365 Fifth Avenue, New York, NY 10016,
USA\\
\vspace{0.3cm} ${}^g$ INFN, Sezione di Padova, Via Marzolo 8, I-35131
Padova, Italy \\
\vspace{0.3cm} ${}^h$ Mani L. Bhaumik Institute for Theoretical
Physics, UCLA Department of Physics and Astronomy, Los Angeles, CA
90095, USA \\
\vspace{0.3cm} ${}^i$ Dipartimento di Fisica e Astronomia,
Universit\`a degli Studi di Padova, Via Marzolo 8, \\I-35131 Padova,
Italy \\
\vspace{0.3cm} ${}^j$ Dipartimento di Matematica e Fisica,
Universit\`a degli Studi Roma Tre, \\ and INFN, Sezione di Roma Tre,
Via della Vasca Navale 84, I-00146 Roma, Italy \\
\vspace{0.3cm} ${}^k$ Max-Planck-Institut f\"ur Physik,
Werner-Heisenberg-Institut,\\ F\"ohringer Ring 6, D-80805 M\"unchen,
Germany \\
\vspace{0.3cm} ${}^l$ Institute for Particle Physics Phenomenology,
University of Durham \\ South Road, Durham DH1 3LE, United Kingdom\\ }
\vspace{0.3cm}
\setcounter{footnote}{0}
\end{center}
\vspace{2ex}

\begin{center}
\begin{minipage}{15.3truecm}
{ We present the first calculation of the complete set of NNLO QED
  corrections for muon-electron scattering. This includes leptonic,
  non-perturbative hadronic, and photonic contributions. All fermionic
  corrections as well as the photonic subset that only corrects the
  electron or the muon line are included with full mass
  dependence. The genuine four-point two-loop topologies are computed
  as an expansion in the small electron mass, taking into account
  both, logarithmically enhanced as well as constant mass effects
  using massification. A fast and stable implementation of the
  numerically delicate real-virtual contribution is achieved by
  combining {\sc OpenLoops} with next-to-soft stabilisation. All
  matrix elements are implemented in the \mcmule{} framework, which
  allows for the fully-differential calculation of any infrared-safe
  observable. This calculation is to be viewed in the context of the
  MUonE experiment requiring a background prediction at the level of
  10 ppm. Our results thus represent a major milestone towards this
  ambitious precision goal.  }
\end{minipage}
\end{center}

\newpage

\input{01_introduction}
\input{02_technical}
\input{03_results}
\input{04_checks}

\input{05_outlook}

\subsection*{Acknowledgement}

We are pleased to acknowledge the Theory Group of the MUonE
initiative/experiment. In particular, Massimo Passera, for his
constant support and stimulating discussions, as well as for
collaboration at early stages; Carlo Carloni, Matteo Fael, Guido
Montagna, Fulvio Piccinini, for interesting discussions and
comparisons, at various stages.

We are grateful to the Mainz Institute of Theoretical Physics (MITP)
of the DFG Cluster of Excellence PRISMA$^+$ (Project ID 39083149) for
its hospitality during the workshop "The Evaluation of the Leading
Hadronic Contribution to the Muon $g-2$: Toward the MUonE Experiment".

The work of A.B.~was supported in part by the ERC Starting Grant REINVENT-714788.
T.E.~and M.R.~acknowledge support from the Swiss National Science Foundation
(SNSF) under grant 200020\_207386.
The work of A.F. was supported in part by the PSC-CUNY Award 64101-00 52.
The work of M.K.M.~is supported by Fellini - Fellowship for Innovation at INFN funded by the European Union's Horizon 2020 research and innovation programme under the Marie Sk{\l}odowska-Curie grant agreement No 754496.
The work of J.R.~is supported by the Italian Ministry of Research (MUR) under grant PRIN 20172LNEEZ.
The work of W.J.T.~received funding from the European Research Council (ERC) under the European Union's Horizon 2020 research and innovation programme (grant agreement No 725110), {\it Novel structures in scattering amplitudes}.
Y.U.~acknowledges support by the UK Science and Technology Facilities Council (STFC) under grant ST/T001011/1.
The work of M.Z.~was supported by the Swiss National Science Foundation (SNSF) under the Ambizione grant PZ00P2-179877.

\bibliographystyle{JHEP}
\bibliography{biblist}

\end{document}

%% file: 01_introduction.tex

\section{Introduction}

The MUonE experiment~\cite{Abbiendi:2016xup, Spedicato:2022qtw,
  Abbiendi:2022oks} aims to measure the differential cross section for
muon-electron elastic scattering by colliding a $160$~GeV muon beam
with atomic electrons located on thin target plates. The purpose is to
derive very precisely the running of the electromagnetic coupling
$\alpha$ at low energy using the method proposed
in~\cite{CarloniCalame:2015obs}.

The need for such a measurement arises from the long-standing tension
between the measured and calculated values of the anomalous magnetic
moment of the muon, $a_\mu = (g -2)/2$. The discrepancy between
$a_\mu$ calculated in the Standard Model~(SM) and the corresponding
experimental measurement was first reported by the BNL E821
experiment~\cite{Muong-2:2006rrc} and recently confirmed by the first
results from the FNAL $g-2$ experiment~\cite{Muong-2:2021ojo}. The
deviation between the combination of the BNL and FNAL experiments, and
the currently accepted SM prediction~\cite{Aoyama:2020ynm} is of
$4.2\sigma$.

It is important to rule out the possibility that the deviation is due
to a systematic error in the calculation. The contribution from the
hadronic vacuum polarisation~(HVP), $a_\mu^{{\tiny \mbox{HVP}}}$,
enters in the SM prediction and cannot be calculated in perturbation
theory. Consequently, this quantity is usually determined from
low-energy electron-positron annihilation data through a dispersive
approach~\cite{Davier:2019can,Keshavarzi:2019abf}. Interestingly, a
recent calculation in lattice QCD~\cite{Borsanyi:2020mff} leads to a
value that is in contrast with the data-driven calculation and reduces
the discrepancy between theory and experiment in the muon magnetic
moment. In light of this situation, it is crucial to pursue new and
different methods to determine the HVP contribution.

The method proposed in~\cite{CarloniCalame:2015obs,Balzani:2021del}
allows for the determination of $a_\mu^{{\tiny \mbox{HVP}}}$ from the
measurement of the running electromagnetic coupling in the space-like
region, which can be carried out by the MUonE experiment. Contrary to
the conventional time-like approach, the space-like region has the
advantage of being smooth and free of hadronic resonances. The
experiment suffers, however, from complications related to the
measurement of a subleading effect. The contribution of the HVP
changes the differential cross section by only up to
$\mathcal{O}(10^{-3})$. As a consequence, the experimental and
theoretical uncertainties should not exceed 10 ppm in order to allow
for a competitive determination of $a_\mu^{{\tiny \mbox{HVP}}}$.

The feasibility of this ambitious precision goal crucially relies on
two special features of muon-electron scattering at the proposed
energy scale. First, the presence of a low-signal normalisation region
allows for the cancellation of experimental systematic
uncertainties. Second, non-perturbative hadronic corrections other
than HVP only enter beyond next-to-next-to-leading order~(NNLO) and
can safely be neglected. This includes the notoriously difficult
hadronic light-by-light scattering contribution. The NNLO HVP
corrections, on the other hand, are significantly simpler and have
been computed in~\cite{Fael:2019nsf} with a dispersive approach. To
avoid any dependence on time-like data the analogous calculation has
been performed with the space-like hyperspherical method
in~\cite{Fael:2018dmz}.

To ensure a clean extraction of the HVP with MUonE, it is further
important to rule out any possible contamination of the signal due to
physics beyond the SM~(BSM). Dedicated studies performed
in~\cite{Masiero:2020vxk,Dev:2020drf} have shown that such effects
could only affect the signal below 10 ppm, if existing BSM bounds are
to be observed. In these analyses the normalisation region was used to
cancel larger effects. This explains the different conclusion reached
in~\cite{Schubert:2019nwm}, where the normalisation region was not
exploited. Even though this latter study is therefore not directly
applicable to the HVP measurement, it opens up the avenue for
dedicated BSM searches with MUonE. This option was further
investigated in~\cite{GrillidiCortona:2022kbq, Galon:2022xcl,
  Asai:2021wzx} where it was shown that new parameter space for light
new physics could indeed be explored.

At present, the main concern on the theoretical side is the
perturbative prediction of the SM background at the level of 10
ppm. It is therefore mandatory to incorporate NNLO corrections in
QED. In addition, it will be necessary to improve the precision of the
calculation by supplementing it with large logarithmic corrections of
soft and collinear origin beyond NNLO. In order to reach this goal,
there has been a coordinated theoretical
effort~\cite{Banerjee:2020tdt} with the goal of developing two
completely independent Monte Carlo event generators.

The {\sc Mesmer}~\cite{CarloniCalame:2020yoz} Monte Carlo is based on
photon-mass regularisation combined with a slicing approach to cope
with soft divergences in the phase-space integration. The complete set
of electroweak corrections are implemented at next-to-leading
order~(NLO) accuracy~\cite{Alacevich:2018vez}. At NNLO, virtual as
well as real leptonic contributions have been included in the form of
leptonic vacuum polarisation and lepton pair
production~\cite{Budassi:2021twh}.  In addition, virtual hadronic
contributions have been included through the time-like
approach. However, NNLO photonic corrections have been calculated
in~\cite{CarloniCalame:2020yoz} employing a
YFS-inspired~\cite{Yennie:1961ad} approximation when dealing with the
genuine two-loop four-point topologies.  Otherwise, the contributions
that only correct the electron or muon line are calculated without any
approximation. Most recently, also the background due to pion
production was studied with {\sc Mesmer} in~\cite{Budassi:2022kqs}.

The second Monte Carlo is being developed within the \mcmule{}
framework~\cite{Banerjee:2020rww} and it is employed in the present
paper. It is based on dimensional regularisation and the FKS$^\ell$
subtraction scheme~\cite{Engel:2019nfw}, a generalisation of the FKS
method~\cite{Frixione:1995ms, Frederix:2009yq} for QED with massive
fermions to any order in perturbation theory. The subset of NNLO QED
corrections restricted to the electron line has been implemented
in~\cite{Banerjee:2020rww}. Perfect agreement was found in a dedicated
comparison with the {\sc Mesmer} code. In the following, we present
the calculation of the full set of NNLO QED corrections in \mcmule{}.

Contrary to QCD experiments, MUonE observables are not collinear safe
and therefore highly sensitive to fermion-mass effects.  It is
therefore not permissible to neglect the electron mass, even though it
is small compared to all other scales in the process. This has
far-reaching ramifications. On the one hand, it simplifies the
infrared (IR) structure since collinear singularities are naturally
regularised by fermion masses.  At the same time, it significantly
complicates the evaluation of loop integrals. In the case of closed
fermion loops this is unproblematic, since semi-numerical methods can
easily be applied. In particular, we have implemented the
aforementioned hyperspherical results from~\cite{Fael:2018dmz} to
calculate leptonic and non-perturbative hadronic NNLO corrections. In
the case of the photonic two-loop amplitude, on the other hand, the
complete calculation with full mass dependence is still not available.

Very recently, however, the analytic evaluation of the amplitude of
di-muon production via massless electron--positron annihilation in
QED~\cite{Bonciani:2021okt}, as well as the heavy-quark pair
production through light-quark annihilation in
QCD~\cite{Mandal:2022vju}, became available, based on the master
integrals computed in~\cite{Mastrolia:2017pfy,
  DiVita:2018nnh,Mandal:2022vju}.  These two results have shown the
feasibility of a completely analytical calculation of the NNLO virtual
muon-electron elastic scattering in the approximation of a massless
electron, presented here for the first time.  This result can be
employed in the context of the MUonE observables if supplemented with
finite electron-mass effects introduced via massification, which was
originally developed in~\cite{Penin:2005eh,Mitov:2006xs,Becher:2007cu}
and later extended to heavy external states
in~\cite{Engel:2018fsb}. It exploits the universal structure of
collinear degrees of freedom to determine the leading mass effects in
a process-independent way. This includes the logarithmically enhanced
as well as the constant terms and only neglects polynomially
suppressed contributions. Since the electron mass is much smaller than
all other scales in the process, this procedure is expected to yield a
reliable approximation of the full mass dependence.

At the same time, this scale hierarchy gives rise to numerical
instabilities in the case of soft and collinear photon
emission. Following~\cite{Banerjee:2021mty} we apply next-to-soft
(NTS) stabilisation combined with {\sc
  OpenLoops}~\cite{Buccioni:2017yxi, Buccioni:2019sur} to ensure a
fast and stable implementation of the delicate real-virtual
amplitude. This method is based on the idea of expanding the
real-virtual contribution in the soft photon energy including the
next-to-leading power term in numerically delicate phase-space
corners. Similar to the eikonal contribution at leading power, also
the subleading term can be related to the non-radiative process. This
was shown in~\cite{Engel:2021ccn} where the Low-Burnett-Kroll
theorem~\cite{Low:1958sn, Burnett:1967km} was extended to the one-loop
level.

The paper is structured as follows. Section~\ref{sec:technical}
presents the technical details of the calculation. This includes a
discussion of the one-loop radiative and the two-loop amplitude, as
well as details on the implementation in the \mcmule{} framework.
Section~\ref{sec:results} presents results for observables that are
relevant for the MUonE experiment.  The calculation has been validated
with various tests, including, among others, a dedicated study for the
reliability of the massified approximation as well as for the NTS
stabilisation, which are discussed in Section~\ref{sec:check}.  We
conclude and remark on future steps in Section~\ref{sec:conclusion}.

%% file: 02_technical.tex
\section{Technical details}
\label{sec:technical}

The complete NNLO results for muon-electron scattering presented in
this article represent the culmination of a program involving various
groups over several years. They are based on numerous concepts and
calculations that have been developed for this program and were
published in separate articles. In this section, we will build on
these articles, give an overview of how the partial results are
combined and describe what additional steps were required. In the
interest of conciseness, we will refrain from repeating the details of
previous work that is required to obtain the physical results
presented here. Instead, we will refer the reader at the appropriate
places to those earlier papers for further details.

\subsection{Overview} \label{sec:techoverview}

The fully differential computation of elastic muon-electron scattering
at higher orders in $\alpha$ follows closely the procedure outlined in
\cite{Banerjee:2020tdt} whose notation we also adopt. To obtain
results at NNLO we need to consider the processes with up to two
additional photons,
\begin{align}
\label{eq:kinematics}
  e^-(p_1)\, \mu^\pm(p_2) \to e^-(p_3)\, \mu^\pm(p_4) +
  \{\gamma(k_1)\, \gamma(k_2)\} \, .
\end{align}
Lepton pair production, i.e. $e^- \mu^\pm\to e^-\mu^\pm (e^+e^-)$,
will not be considered here. Even though this process plays an
important role in a future MUonE analysis, it is a measurably
different physical process and from a theoretical point of view can be
considered in isolation \cite{Budassi:2021twh}.  As discussed in
\cite{Banerjee:2020tdt}, the trivial inclusion of the tree-level $Z$
exchange is needed to match the precision required by MUonE, but the
full NLO electroweak effects are below the 10\,ppm target precision as
shown in a calculation by {\sc Mesmer}~\cite{Alacevich:2018vez}.  We
have implemented the same corrections in {\sc McMule} in the context
of $ee\to\mu\mu$~\cite{Kollatzsch:2022bqa} and found full agreement.
Nevertheless, these corrections are not considered in this paper where
we restrict ourselves to pure QED.

We will denote the $\ell$-loop amplitude of the process
\eqref{eq:kinematics} with $j$ additional photons by
$\Amp{n+j}{\ell}$. The (differential) LO cross section $\D\sigma_0$ is
obtained by integrating the tree-level (squared) matrix element for
the $2\to 2$ process $\M{n}{0}\equiv|\Amp{n}{0}|^2$ over the
two-particle phase space $\D\Phi_n$.

In addition to the flux factor, a measurement function
$O(p_3,p_4,\{k_1,k_2\})$ that defines the observable is implicitly
understood to be part of the phase-space integration. The only
constraint on this function is that it represents an IR-safe
quantity. This property is crucial to ensure the cancellation of IR
singularities when going beyond LO. Concretely, in the soft limit of
e.g.~$k_1\to 0$ we require $O(p_3,p_4,k_1,\{k_2\}) \to
O(p_3,p_4,\{k_2\})$.  Contrary to computations in QCD, there is no
requirement on $O$ for a collinear limit $k_i\|p_j$. The potential
non-cancellation of contributions from collinear regions of the phase
space is regularised by the fermion masses. The corresponding
logarithms are physical.  They can be and are measured. As a
consequence, in order to be fully differential, in particular with
respect to emission of collinear photons, muon mass $M$ and electron
mass $m$ effects have to be included.

As we will discuss below, our calculation includes full dependence on
$M$, but makes certain approximations regarding the $m$
dependence. These approximations are restricted to the (finite part of
the) two-loop matrix elements $\cM_n^{(2)}$, as given in
\eqref{AMPvv}, and are driven by the complexity of two-loop
calculations with many scales. Even with these approximations, there
are no collinear singularities present and only soft singularities
appear.  This leads to a tremendous simplification of the IR structure
and allows the use of the FKS$^{2}$ subtraction
method~\cite{Engel:2019nfw} to perform a fully differential
phase-space integration also at NNLO.

Radiative corrections to muon-electron scattering have been considered
in the past~\cite{Nikishov:1961, Eriksson:1961, Eriksson:1963,
  VanNieuwenhuizen:1971yn, Kukhto:1987uj, Bardin:1997nc,
  Kaiser:2010zz}, using various approximations.  Two independent fully
differential results with complete muon and electron mass dependence
exist and have been
compared~\cite{Alacevich:2018vez,Banerjee:2020rww}. As is well known,
the NLO correction to the (differential) cross section
$\D\sigma^{(1)}$ is a combination of real and virtual contributions
\begin{align}
  \label{Snlo}
   \D\sigma^{(1)} &= \D\sigma^{(\mathrm{v})} + \D\sigma^{(\mathrm{r})}
   = \int \D\Phi_n \, \M{n}{1} + \int \D\Phi_{n+1} \, \M{n+1}{0} \ ,
\end{align}
where $\M{n}{1} = 2\,\Re [\Amp{n}{1}\times (\Amp{n}{0})^*]$ and
$\M{n+1}{0}=|\Amp{n+1}{0}|^2$. For an IR-safe observable, the IR
singularities cancel between $\D\sigma^{(\mathrm{v})}$ and
$\D\sigma^{(\mathrm{r})}$. We do the renormalisation in terms of
on-shell fermion masses and the on-shell coupling
$\alpha\equiv\alpha(0)$. Intermediate results depend on the
regularisation scheme, but this scheme dependence cancels for physical
predictions. In \mcmule{} the standard choice is to use the
four-dimensional helicity scheme ({\sc fdh})~\cite{Gnendiger:2017pys}.

The NNLO corrections to the cross section are obtained as the sum of
three contributions, dubbed double-virtual $({\rm vv})$, real-virtual
$({\rm rv})$, double-real $({\rm rr})$ terms, as
\begin{align}
  \label{Snnlo}
  \D\sigma^{(2)} &= \D\sigma^{(\mathrm{vv})} + \D\sigma^{(\mathrm{rv})}
   + \D\sigma^{(\mathrm{rr})}
    =  \int \D\Phi_n\, \M{n}{2} + \int \D\Phi_{n+1}\, \M{n+1}{1}
     + \int \D\Phi_{n+2}\, \M{n+2}{0}\, ,
\end{align}
where the matrix elements are defined as
\begin{align}
\label{AMPvv}
  \M{n}{2} &= 2\,\Re\Big[ \Amp{n}{2} \times (\Amp{n}{0})^* \big]
           +  \big| \Amp{n}{1} \big|^2
           \equiv \M{n}{2+0} + \M{n}{1+1}\, ,
\\[3pt]
\label{AMPvr}
  \M{n+1}{1} &=  2\,\Re\big[ \Amp{n+1}{1}\times (\Amp{n+1}{0})^* \big] \, ,
\\[3pt]
\label{AMPrr}
  \M{n+2}{0} &= \big| \Amp{n+2}{0} \big|^2 \, .
\end{align}
Taken separately, the three parts of \eqref{Snnlo} are IR
divergent. Following the FKS$^2$ procedure~\cite{Engel:2019nfw}, we
re-express the NNLO corrections as a sum of three separately finite
parts
\begin{align}
  \label{FKSnnlo}
  \D\sigma^{(2)} &= \D\sigma_{n}^{(2)}(\xi_c) + \D\sigma_{n+1}^{(2)}(\xi_c)
  + \D\sigma_{n+2}^{(2)}(\xi_c) \, .
\end{align}
According to \eqref{FKSnnlo}, $\D\sigma^{(2)}$ consists of a
two-particle part $\D\sigma_{n}^{(2)}(\xi_c)$ that combines the double
virtual with suitable integrated real corrections to obtain an IR
finite result, a three-particle part $\D\sigma_{n+1}^{(2)}(\xi_c)$
related to real-virtual corrections, and a four-particle part
$\D\sigma_{n+2}^{(2)}(\xi_c)$ that corresponds to soft-subtracted
double-real corrections. These parts individually depend on an
unphysical parameter $\xi_c$. As will be illustrated in
Section~\ref{sec:impl}, this $\xi_c$ dependence has to cancel in the
sum, providing a very useful check of a correct and numerically stable
implementation.

\begin{figure}[t]
\centering
  \includegraphics[width=0.3\textwidth]{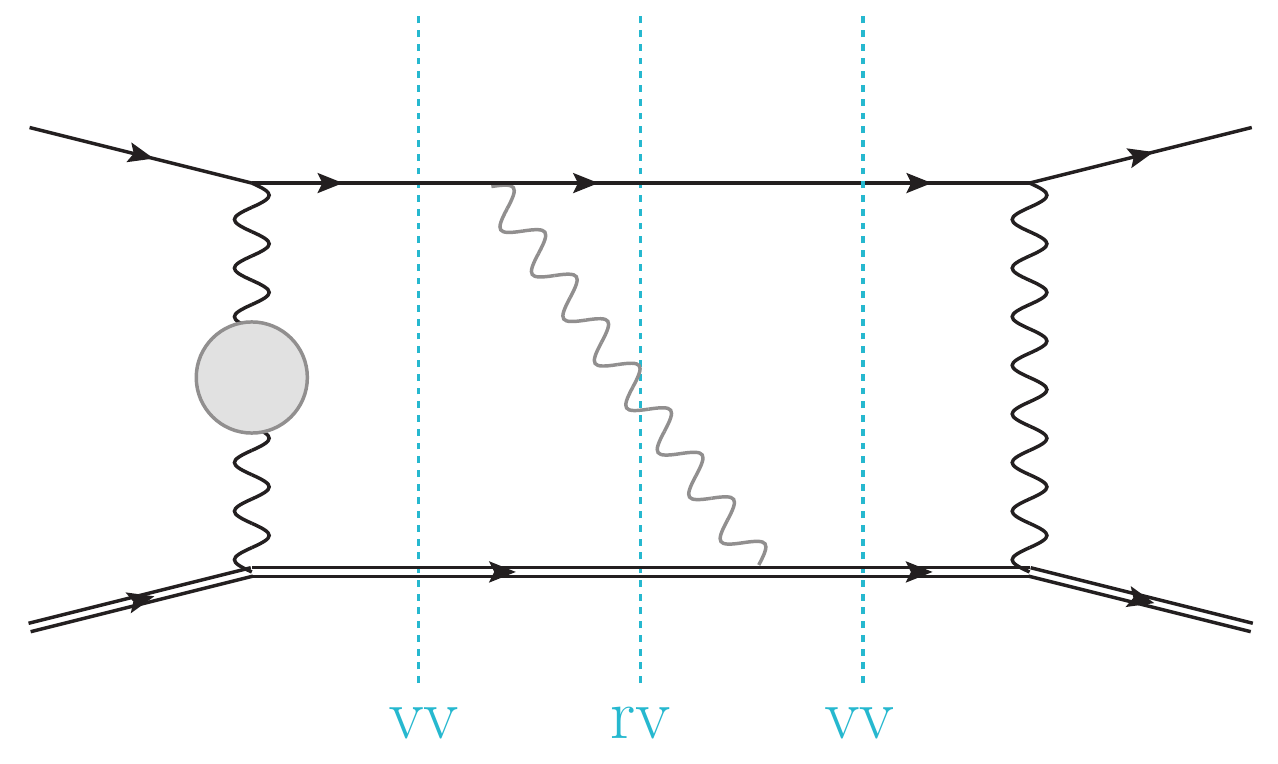}
\quad
  \includegraphics[width=0.3\textwidth]{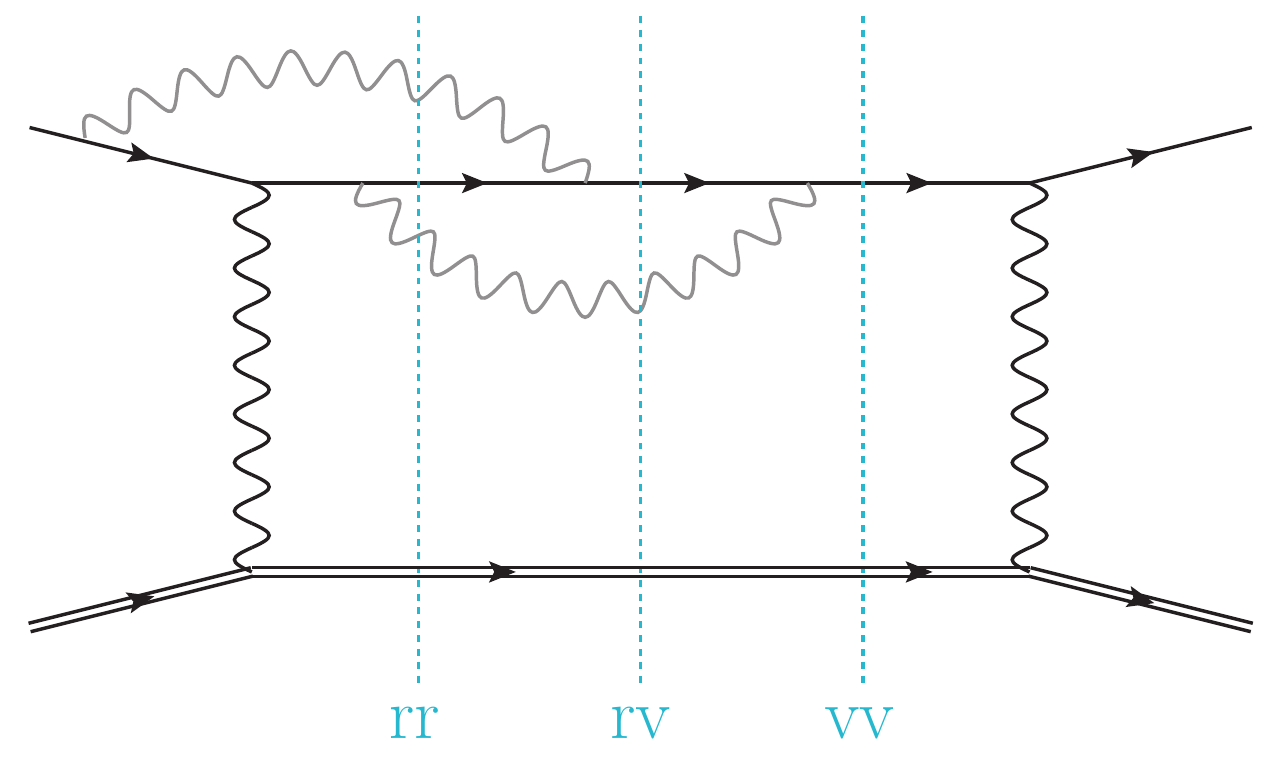}
\quad
  \includegraphics[width=0.3\textwidth]{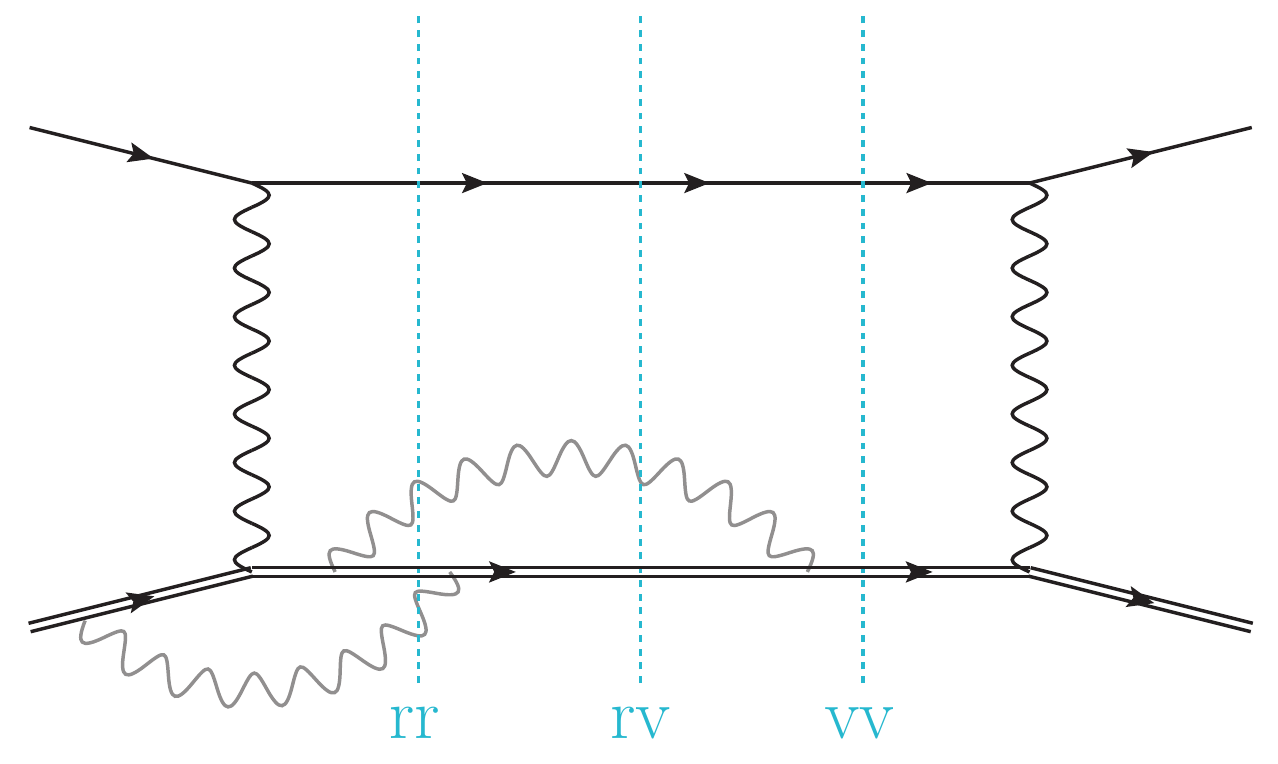}
\\[10pt]
  \includegraphics[width=0.3\textwidth]{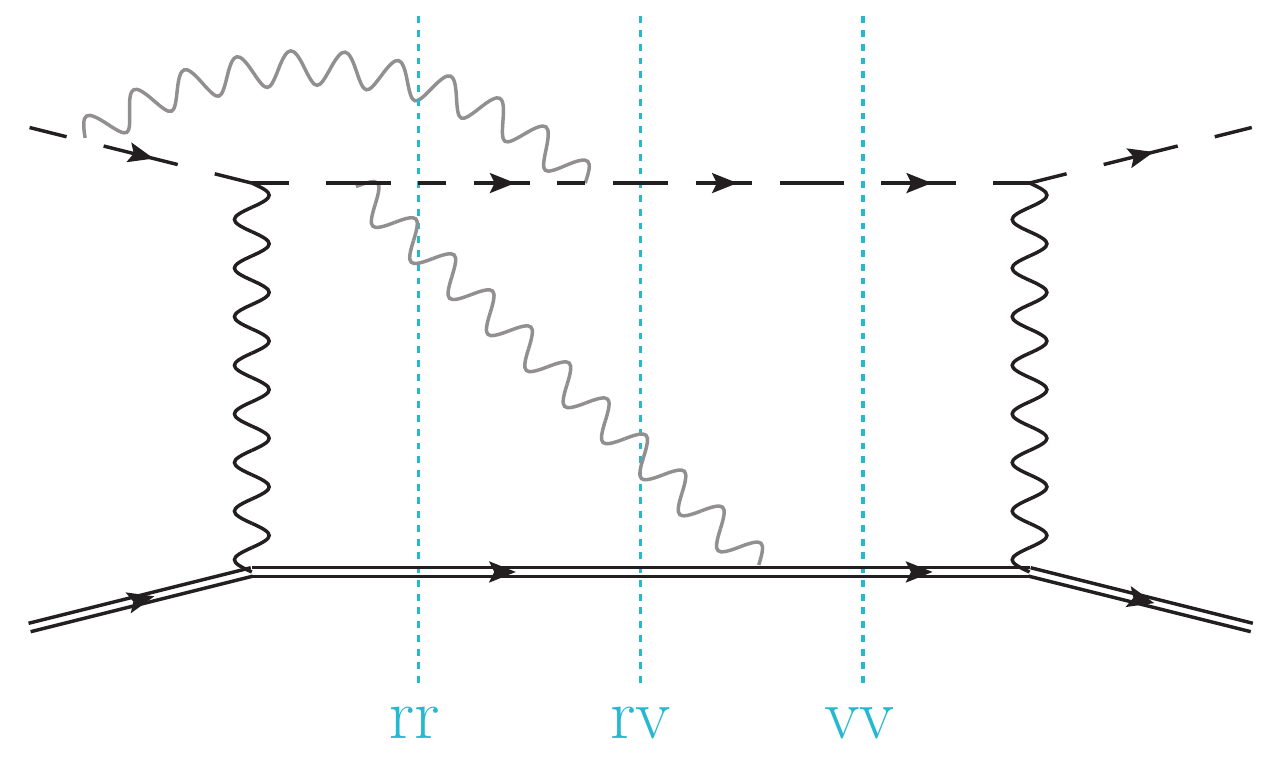}
\quad
  \includegraphics[width=0.3\textwidth]{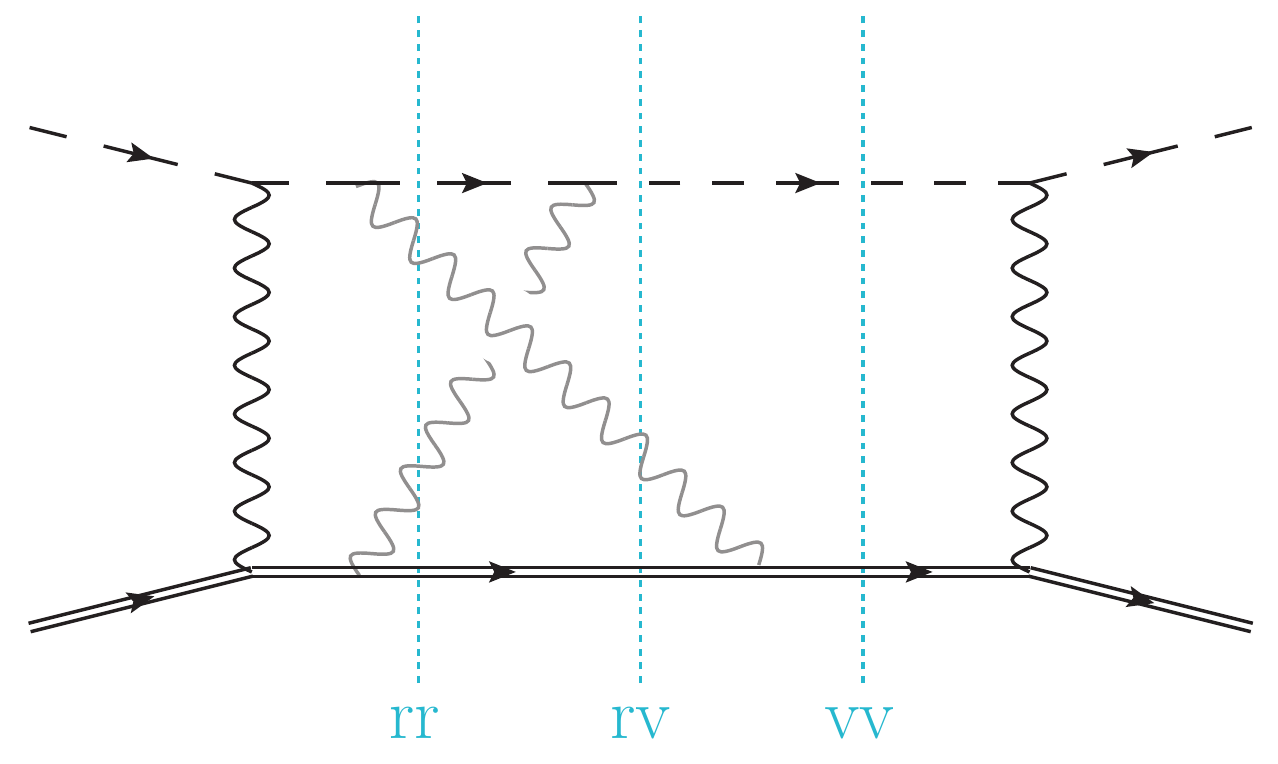}
\quad
  \includegraphics[width=0.3\textwidth]{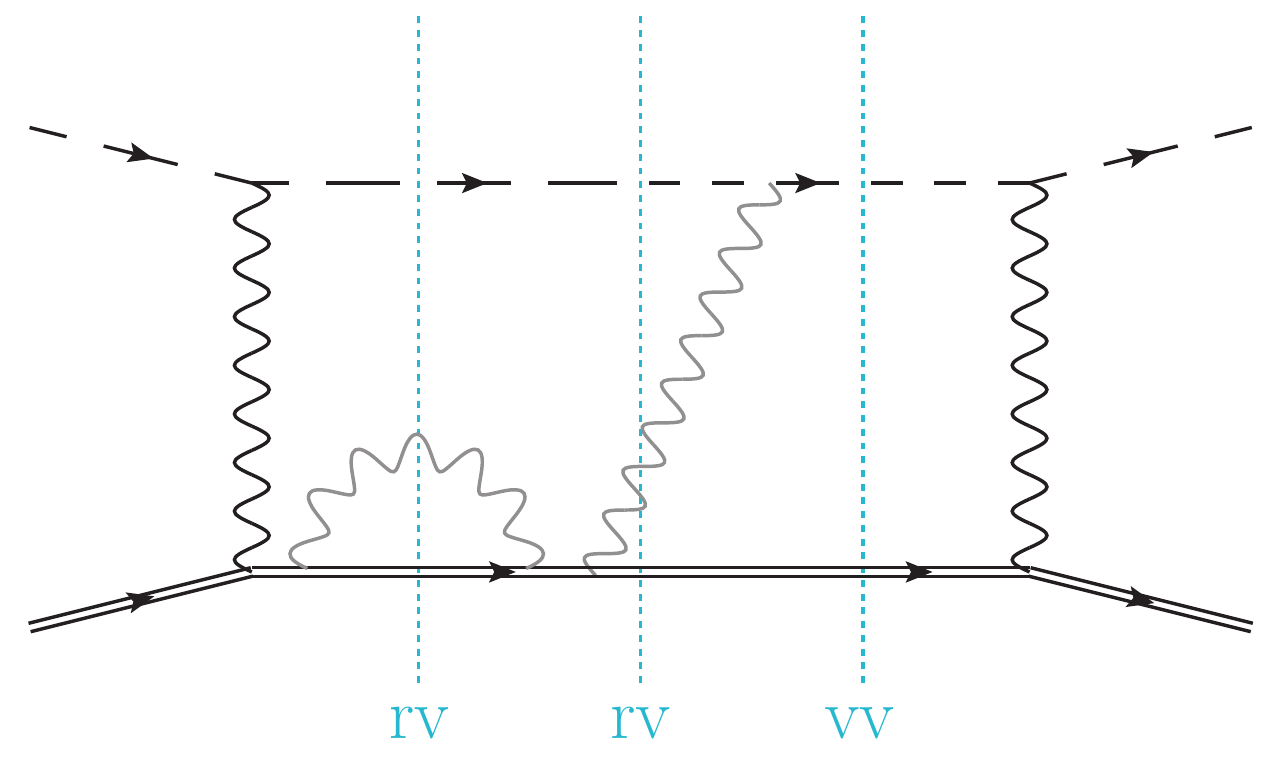}
\caption{\label{fig:fd} Sample contributions to the (squared) matrix
  elements $\cM_{n}^{(2)}$ (vv), $\cM_{n+1}^{(1)}$ (rv), and
  $\cM_{n+2}^{(0)}$ (rr). The leading order part $\cM_{n}^{(0)}\sim
  q^2\,Q^2$ is shown in black while additional particles are shown in
  grey. The top-left panel shows fermionic corrections. Electronic
  (muonic) corrections are shown in the top-middle (right) panel, with
  all additional emissions $\sim q^4$ ($\sim Q^4$) from the electron
  (muon) line. Muon mass effects (solid double line) and electron mass
  effects (single solid line) are taken into account exactly. The
  bottom panel shows examples of mixed corrections, with additional
  couplings $q^3\,Q$ (left panel), $q^2\,Q^2$ (middle panel), and
  $q\,Q^3$ (right panel). For these contributions electron mass
  effects in $\cM_{n}^{(2)}$ (vv) are included only through
  massification (see text), as indicated by the dashed single line.  }
\end{figure}

The higher-order corrections to the cross section can be split into
photonic and fermionic corrections. The latter contain leptonic vacuum
polarisation (VP) contributions, $\D\sigma_{\mathrm{lep}}^{(i)}$, and
are separately gauge independent. We include electron, muon, and tau
loops in our results with all mass effects. In addition we also
provide results for hadronic loops, $\D\sigma_{\mathrm{had}}^{(i)}$,
i.e. the signal of the MUonE experiment. The photonic corrections can
be split further into separately gauge independent pieces by formally
differentiating between the electron charge $q$ and the muon charge
$Q$, and organising the matrix elements according to powers of $q$ and
$Q$. The tree-level matrix element of the $2\to 2$ process has
couplings $\cM_n^{(0)} \sim q^2\, Q^2$.  At NLO, there are
contributions with an additional factor $q^2$ (electronic corrections
$\D\sigma_e^{(1)}$), an additional factor $Q^2$ (muonic corrections
$\D\sigma_\mu^{(1)}$), and an additional factor $q\,Q$ (mixed
corrections $\D\sigma_{e\mu}^{(1)}$). In analogy to lepton-proton
scattering, the latter are known as two-photon exchange contributions.
With respect to the Born contribution, at NNLO there are four
additional factors of $q$ or $Q$ for the photonic
corrections. Following the NLO terminology, we will consider a split
into electronic ($\D\sigma_e^{(2)}$, additional factor $q^4$), muonic
($\D\sigma_\mu^{(2)}$, additional factor $Q^4$) and various mixed
corrections with additional factors ($q^3Q$, $q^2Q^2$, $qQ^3$).  The
latter are often combined into $\D\sigma_{e\mu}^{(2)}$ for the
presentation of the results, even though they are computed
separately. Hence, at NLO ($\ell=1$) and NNLO ($\ell=2$), we decompose
\eqref{Snlo} and \eqref{Snnlo} as
\begin{align}
  \label{SGI}
   \D\sigma^{(\ell)} &= \D\sigma_e^{(\ell)} + \D\sigma_{e\mu}^{(\ell)} +
   \D\sigma_\mu^{(\ell)} + \D\sigma_{\mathrm{lep}}^{(\ell)} +
   \D\sigma_{\mathrm{had}}^{(\ell)}\, .
\end{align}
Some (squared) matrix elements contributing to $\D\sigma^{(2)}$ are
depicted in Figure~\ref{fig:fd}. An example for the fermionic
corrections $\D\sigma_{\mathrm{lep}}^{(\ell)}$ and
$\D\sigma_{\mathrm{had}}^{(\ell)}$ is shown in the top left panel of
Figure~\ref{fig:fd}. The vv cuts correspond to $\cM_n^{(2)}$, the
first due to the one-loop amplitude squared $\M{n}{1+1}$, the second
due to the interference of the two-loop amplitude with the tree-level
amplitude $\M{n}{2+0}$. The rv cut corresponds to a $\cM_{n+1}^{(1)}$
contribution. For photonic corrections we also have rr cuts, due to
double real contributions involving $\cM_{n+2}^{(0)}$. Contributions
to $\D\sigma_e^{(2)}$ and $\D\sigma_\mu^{(2)}$ are also shown in the
top panel of Figure~\ref{fig:fd}, whereas the bottom panel depicts
various contributions to $\D\sigma_{e\mu}^{(2)}$.

The split \eqref{SGI} has a threefold motivation. First, since the
electron mass is much smaller than the muon mass, the electronic
corrections are expected to be numerically more important. As we will
see, however, this expectation is only partially correct. Second, the
electronic and muonic corrections can be computed with full dependence
on the electron mass $m$ as well as the muon mass
$M$~\cite{CarloniCalame:2020yoz, Banerjee:2020rww}, since the two-loop
amplitudes can be expressed through the massive two-loop form
factor~\cite{Mastrolia:2003yz,Bonciani:2003ai,Bernreuther:2004ih}. We
use a solid single line for the electron in Figure~\ref{fig:fd} to
indicate that all $m$ effects are fully taken into account. As we will
see below, for the mixed corrections shown in the bottom panel of
Figure~\ref{fig:fd}, electron mass effects in the two-loop matrix
element are taken into account approximately, as indicated by using a
dashed single line for the electron. Finally, changing from $\mu^+$ to
$\mu^-$ simply amounts to the replacement $Q\to -Q$.  In particular,
the electronic and muonic corrections are not affected.

Once the amplitudes for muon-electron scattering are known, we can
follow the procedure used previously for Bhabha
scattering~\cite{Banerjee:2021mty}, M{\o}ller
scattering~\cite{Banerjee:2021qvi}, and lepton pair
production~\cite{Kollatzsch:2022bqa} to obtain physical results. The
double radiative tree-level matrix element $\cM_{n+2}^{(0)}$ is
trivial to compute with full $M$ and $m$ dependence. The only delicate
point is to ensure a numerically reliable integration over those
phase-space regions that lead to collinear pseudo-singularities. This
is achieved by a partitioning and tuning of the phase space to have a
direct match of the small angle with an integration
variable~\cite{Engel:2022kde}. As mentioned above, the check of the
independence on $\xi_c$ of the final result plays an important role
here. The remaining matrix elements, $\cM_{n+1}^{(1)}$ and
$\cM_n^{(2)}$, are more delicate and will be discussed below, in
Sections~\ref{sec:OLrad} and \ref{sec:TL}.

\subsection{One-loop radiative matrix element} \label{sec:OLrad}

The real-virtual matrix element \eqref{AMPvr} is particularly delicate
with respect to its numerical stability. For the bulk of the phase
space this contribution was computed with {\sc
  OpenLoops}~\cite{Buccioni:2017yxi,Buccioni:2019sur}, which
recursively constructs amplitudes from process-independent building
blocks given by the Feynman rules of the model. Recently, {\sc
  OpenLoops} was extended to allow for the separate calculation of
QCD, QED and weak corrections with variable number of leptons and
quarks, as well as with massive leptons. In order to comply with the
split of the photonic corrections discussed above, we separated the
purely muonic, purely electronic and mixed corrections through a power
counting in the muon and electron charges $Q$ and $q$.

The {\sc OpenLoops} program provides high numerical stability in
one-loop amplitudes due to the on-the-fly reduction of tensor
integrals\footnote{ The remaining scalar two-, three- and four-point
  integrals are evaluated with the external tools Collier
  \cite{Denner:2016kdg} (double precision calculation) or OneLoop
  \cite{vanHameren:2010cp} (quadruple precision).}
\cite{Buccioni:2017yxi}, dedicated kinematic expansions, and a hybrid
precision mode \cite{Buccioni:2019sur}. In the latter, the majority of
the {\sc OpenLoops} recursion steps is performed in double precision,
while only the most critical steps of the on-the-fly algorithm are
performed in quadruple precision. While this procedure provides
excellent CPU efficiency and numerical stability for a wide range of
processes, even in ultra-soft and collinear regions
\cite{Buccioni:2019sur}, for simultaneously extremely soft and
collinear kinematics the numerical stability in the hybrid precision
configuration was not sufficient for this process. Hence a
next-to-soft (NTS) stabilisation was employed.

The basic idea of NTS stabilisation~\cite{Banerjee:2021mty} is to
replace the full one-loop matrix elements by a numerically adequate
expression in the critical soft regions of phase space. The well-known
soft limit, given by an eikonal factor times a reduced matrix element,
is not sufficiently accurate.  To improve the situation, the NTS limit
has to be used. This limit can be obtained by a generalisation of the
Low-Burnett-Kroll theorem~\cite{Low:1958sn,Burnett:1967km} to one
loop. In~\cite{Engel:2021ccn} it was shown how to write the NTS limit
for one-loop matrix elements in a process independent way. With these
expressions we can obtain a numerically reliable evaluation of the
one-loop contribution in all regions of phase space. As will be
discussed in Section~\ref{sec:nts}, this implementation was
successfully checked against a full quadruple-precision calculation in
{\sc OpenLoops}.

While NTS stabilisation is crucial for the mixed real-virtual
corrections, it is not strictly necessary in the case of the
electronic and muonic contributions. For the results presented
in~\cite{Banerjee:2020rww}, we have used an in-house calculation of
the real- virtual contribution assisted by
Collier~\cite{Denner:2016kdg} in problematic regions of the phase
space. However, this implementation is superseded in the current
version of the code by {\sc OpenLoops} combined with NTS
stabilisation. Even though this results in a slower evaluation speed
for the matrix element, the improved numerical stability ensures a
faster and more reliable phase-space integration.

\subsection{Double-virtual matrix element} \label{sec:TL}

As a final ingredient for the NNLO corrections, we need the
double-virtual matrix element $\cM_n^{(2)}$.

Following~\cite{Fael:2018dmz}, the fermionic contributions at two
loop, $\D\sigma_{\mathrm{lep}}^{(2)}$ and
$\D\sigma_{\mathrm{had}}^{(2)}$, can be calculated with full $M$ and
$m$ dependence, using the hyperspherical
method~\cite{Levine:1974xh,Levine:1975jz,Laporta:1994mb}. This
semi-numerical approach is independent of the exact form of the
VP. This therefore makes it possible to compute leptonic as well as
non-perturbative hadronic contributions simultaneously. In the case of
the leptonic VP the analytic two-loop result
from~\cite{Djouadi:1993ss} can be used. For the HVP, on the other
hand, we can rely on the Fortran library
\texttt{alphaQED}~\cite{alphaqed}.

For the photonic part of $\cM_n^{(2)}$, according to \eqref{AMPvv}, we
need $\Amp{n}{1}$ and $\Amp{n}{2}$, decomposed as
\begin{align}
\label{AMPnlo}
\Amp{n}{1} &=
  q^3 Q\, \A311
+ q^2 Q^2\, \A221
+ q Q^3\, \A131
\ , \\
\label{AMPnnlo}
\Amp{n}{2} &= q^5 Q\, \A512
  + q^4 Q^2\, \A422
  + q^3 Q^3\,  \A332
  + q^2 Q^4\, \A242
  + q Q^5\, \A152 \, .
\end{align}
The purely electronic and muonic two-loop corrections, $\A512$ and
$\A152$, with complete mass dependence, can be calculated using the
known analytic results of the massive form
factors~\cite{Mastrolia:2003yz,Bonciani:2003ai,Bernreuther:2004ih}.
Therefore, the terms $\D\sigma_e^{(2)}$ and $\D\sigma_\mu^{(2)}$,
having complete dependence on the muon and electron mass, can be
computed within the \mcmule{} framework, as previously employed
in~\cite{Banerjee:2020rww,CarloniCalame:2020yoz}.  On the other hand,
$\D\sigma_{e\mu}^{(2)}$ receives contributions from the remaining
photonic mixed terms, $\A422 \, , \A332 \, , $ and $ \A242 \,$, whose
expressions with full mass dependence are currently not available.
However, the electron mass $m$ is sufficiently small with respect to
all other scales $S$ in the process to approximate
$\D\sigma_{e\mu}^{(2)}$ by neglecting terms that are polynomially
suppressed in $m^2/S$. As explained in detail in
Section~\ref{sec:massif-th}, this expansion can be efficiently
computed based on the massless result using massification.

In the following, we provide the details of the evaluation of the two
terms contributing to $\cM_n^{(2)}$ in \eqref{AMPvv}, namely the
interference term, $\cM_n^{(2+0)} \equiv 2\,\Re\Big[ \Amp{n}{2} \times
  (\Amp{n}{0})^* \big] $, that receives contributions from two-loop
graphs, and the contribution due to squared one-loop graphs,
$\cM_n^{(1+1)} \equiv \big| \Amp{n}{1} \big|^2 $.

\subsubsection{Two-loop contribution}

The analytic evaluation of $\cM_n^{(2+0)}(m=0)$ is carried out by
considering the electron as a massless particle, while retaining full
dependence on the muon mass $M$, and it constitutes one of the main
novel results of this paper.

By adopting the same strategy as used for the crossing-related process
$e^+ e^- \to \mu^+ \mu^-$ in~\cite{Bonciani:2021okt},
$\Amp{n}{2}(m=0)$ has been generated by FeynArts~\cite{Hahn:2000kx}
and FeynCalc~\cite{Shtabovenko:2016sxi}, within conventional
dimensional regularisation ({\sc cdr}). The decomposition of
$\cM_n^{(2+0)}$ has been carried out by the in-house calculation
framework {\sc aida}~\cite{Mastrolia:2019aid}, that combines the
Adaptive Integrand Decomposition~\cite{Mastrolia:2016dhn,
  Mastrolia:2016xxx}, and the integration-by-parts decomposition,
through the interfaces to Reduze~\cite{vonManteuffel:2012np}
libraries.  The master integrals for the muon-electron elastic
scattering were computed analytically in~\cite{Mastrolia:2017pfy,
  DiVita:2018nnh, Mandal:2022vju} by means of the method of
differential equations~\cite{Barucchi:1973zm, Kotikov:1990kg,
  Remiddi:1997ny, Gehrmann:1999as, Henn:2013pwa} and the Magnus
exponential~\cite{Argeri:2014qva, DiVita:2014pza}.

The analytic expression is obtained in the non-physical region $s <
0$, $t < 0$, and cast as a Laurent series expansion around $d=4$
space-time dimensions, whose coefficients are a combination of
generalised polylogarithms~(GPLs)~\cite{Goncharov:1998kja,
  Gehrmann:2001jv}, depending on the kinematic variables.  The
numerical values of the amplitude in the physical region of the
elastic scattering, $ s > M^2 $ and $ - (s-M^2)^2/s < t < 0$, can be
obtained by evaluating the GPLs according to the prescription $ s \to
s + i \delta \,$, namely by assigning a small positive imaginary part
$\delta \ll 1$.

Let us remark that the result for the amplitude of muon-electron
elastic scattering is found to obey the crossing symmetry linking it
to di-muon production in electron-positron
fusion~\cite{Bonciani:2021okt}.  Moreover, the diagrams considered in
the current process have been recently used also to derive the
analytic expressions of the colour-stripped partial amplitudes of $q
{\bar q} \to t {\bar t}$ in QCD~\cite{Mandal:2022vju}, proportional to
the colour coefficients corresponding to Abelian-like diagrams, and
were found in agreement with the previously known
results~\cite{Czakon:2008zk, Bonciani:2008az, Bonciani:2009nb}.  The
matrix element has been computed in {\sc cdr}, and it is expressed in
terms of the on-shell muon mass and the QED coupling renormalised in
the $\overline{\text{MS}}$ scheme~\cite{Bonciani:2021okt}.

In view of the electron massification procedure, discussed in
Section~\ref{sec:massif-th}, the fermionic terms coming from diagrams
with closed fermion loops, as well as the terms proportional to
$q^6\,Q^2$ (purely electronic corrections) and $q^2\,Q^6$ (purely
muonic corrections), are excluded from $\cM_n^{(2+0)}(m=0)$, see
\eqref{AMPnnlo}, and replaced by their complete massive version.

\subsubsection{Squared one-loop contribution}

The determination of $\cM_n^{(1+1)}(m=0)$ requires the knowledge of
the squared one-loop amplitude $\Amp{n}{1}(m=0)$ up to
$\mathcal{O}(\epsilon^2)$.  We have performed two independent analytic
calculations of the squared one-loop contribution: one using
FeynArts~\cite{Hahn:2000kx} and FeynCalc~\cite{Shtabovenko:2016sxi} in
{\sc cdr}, and one using QGraf~\cite{Nogueira:1991ex},
Package-X~\cite{Patel:2015tea}, and its companion tool PVReduce in
{\sc fdh}.  In both cases, the intermediate expressions contain only
one-loop scalar integrals that, upon integration-by-parts reduction,
are decomposed in terms of master integrals. The latter have been
evaluated using the same technique as the two-loop integrals mentioned
earlier, i.e.~via the method of differential
equations~\cite{Barucchi:1973zm, Kotikov:1990kg, Remiddi:1997ny,
  Gehrmann:1999as, Henn:2013pwa} and the Magnus
exponential~\cite{Argeri:2014qva,DiVita:2014pza}.  Because of products
of master integrals containing $1/\epsilon^2$ poles, we had to extend
up to $\mathcal{O}(\epsilon^2)$ the one-loop integrals given
in~\cite{Mastrolia:2017pfy}.  The complete expression of the squared
one-loop contribution is finally split into electronic, muonic and
photonic corrections, according to Section~\ref{sec:techoverview}. As
for $\cM_n^{(2+0)}(m=0)$, we also replace the terms $q^2 Q^6$ and $q^6
Q^2$ in $\cM_n^{(1+1)}(m=0)$ by the analogous expression with full $m$
dependence.

\subsubsection{Infrared structure}

As an additional check of $\cM_n^{(2)}(m=0)$, we have verified that it
has the expected IR structure. The residual IR poles present in the
two-loop corrections evaluated at $m=0$ can be obtained by adapting a
procedure originally developed for the IR structure of QCD amplitudes
in~\cite{Becher:2009kw, Becher:2009qa, Hill:2016gdf}. The IR poles are
dictated by an anomalous dimension $\Gamma$.  The explicit expressions
for the perturbative coefficients of $\Gamma$ in {\sc cdr} for the
process $e^+\,e^-\to\mu^+\,\mu^-$ up to $\alpha^2$ can be found in the
supplemental material of~\cite{Bonciani:2021okt}. For our process we
have to use the crossed expressions with $s\leftrightarrow t$.
Through the perturbative coefficients of $\Gamma$ and the QED beta
function, an IR renormalisation factor
$Z_{\text{IR}}=1+Z_{\text{IR}}^{(1)}+Z_{\text{IR}}^{(2)}+\mathcal{O}(\alpha^3)$
can be written. The IR poles remaining in the UV-renormalised virtual
corrections to muon-electron scattering at $\ell$ loops can then be
obtained by multiplying $Z_{\text{IR}}$ by the amplitude up to
($\ell-1$) loops and subsequently collecting the terms proportional to
order $\alpha^\ell$ in the product. In particular, at two loops one
finds
\begin{align} \label{eq:polesM}
  \left. \mathcal{M}^{(2)}_n \right|_{\tiny{\text{poles}}} = & \quad
   \left[2 \left( Z_{\text{IR}}^{(2)} -
    \left(Z_{\text{IR}}^{(1)} \right)^2 \right) \mathcal{M}^{(0)}_n
    + \, Z_{\text{IR}}^{(1)}  \mathcal{M}^{(1)}_n \right]_{\tiny{\text{poles}}}
  \,
  \nonumber \\
  & + \left[Z_{\text{IR}}^{(1)} \mathcal{M}^{(1)}_n
    - \left(Z_{\text{IR}}^{(1)}\right)^2 \mathcal{M}^{(0)}_n \right]_{\tiny{\text{poles}}}\, .
\end{align}
In \eqref{eq:polesM} only the IR poles in the dimensional regulator
have to be retained in both the l.h.s.~and r.h.s.~of the
equations. The first line of \eqref{eq:polesM} leads to the IR poles
of the interference of the two-loop and tree-level amplitudes
$\cM_n^{(2+0)}(m=0)$, while the second accounts for the IR poles
originating from the absolute square of the one-loop amplitude
$\cM_n^{(1+1)}(m=0)$.

\subsubsection{Scheme conversion}

\begin{figure}[t]
    \centering
    \includegraphics[width=0.9\textwidth]{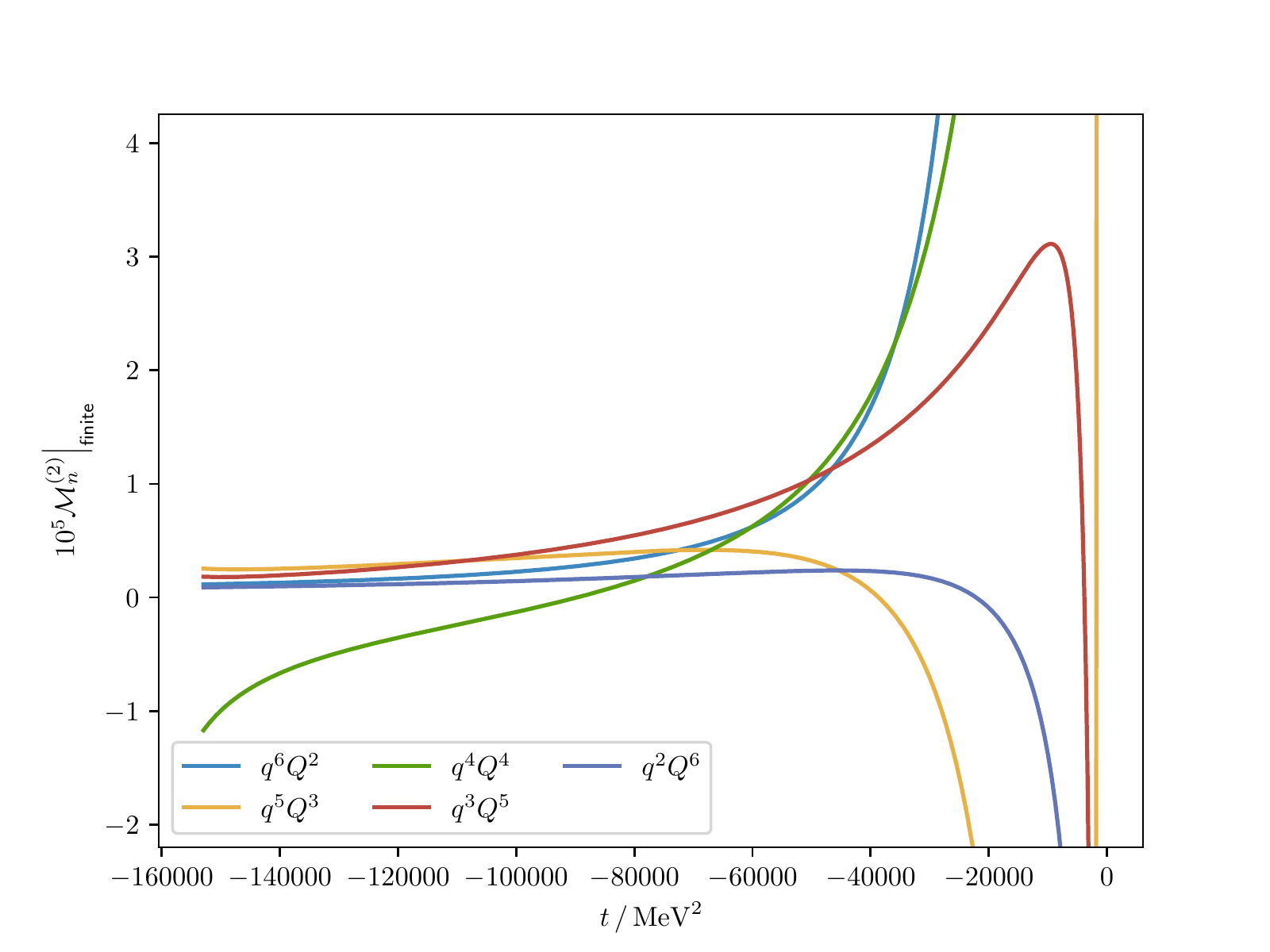}
    \caption{\label{fig:vv} The finite part of the matrix element
      $\mathcal{M}_{n}^{(2)}(m=0)$, defined in~\eqref{AMPvv},
      calculated in the {\sc fdh} scheme, within the kinematic region
      used in Section~\ref{sec:results}, i.e.~$s=174\,684\,{\rm
        MeV}^2$ and $t \in [-153\,069, -1\,021]\,{\rm MeV}^2$.  }
\end{figure}

The massless two-loop amplitude as discussed so far, has been
evaluated in terms of the QED coupling renormalised in the
$\overline{\text{MS}}$ scheme, whereas for the final results we use
the on-shell $\alpha$. Thus, in principle we have to perform a
renormalisation-scheme conversion for $\cM_{n}^{(2)}$. However, the
difference between the two renormalisation schemes is restricted to
fermion loop contributions. As discussed above, the latter are
computed independently and directly in the on-shell scheme. Hence, for
the photonic terms, no further shift is required.

Another scheme conversion that is needed is a regularisation-scheme
change. Following the standard convention of \mcmule{}, the matrix
elements are converted from the {\sc cdr} scheme to the {\sc fdh}
scheme.  This can be done by realising that the IR-subtracted matrix
element is independent of the regularisation
scheme~\cite{Broggio:2015dga, Gnendiger:2016cpg}, i.e.
\begin{align} \label{eq:scheme}
 \Big(Z^\textsc{fdh}_\text{IR}\Big)^{-1} \mathcal{M}_n^{(\textsc{fdh})}
 =
\Big(Z^\textsc{cdr}_\text{IR}\Big)^{-1} \mathcal{M}_n^{(\textsc{cdr})} + \mathcal{O}(\epsilon)\,.
\end{align}
The difference between $Z^\textsc{cdr}_\text{IR}$ and
$Z^\textsc{fdh}_\text{IR}$ is in the perturbative coefficients of
$\Gamma$ and the beta function~\cite{Kilgore:2012tb,
  Gnendiger:2014nxa, Broggio:2015dga, Gnendiger:2016cpg}.  However,
the structure of $Z_{\text{IR}}$ and the basic ideas remain the
same. With $Z^\textsc{fdh}_\text{IR}$ known, we repeat the calculation
of the one-loop matrix element in {\sc fdh} and can obtain
$\mathcal{M}_n^{(2)}$ in the {\sc fdh} scheme, using
\eqref{eq:scheme}.

In Figure~\ref{fig:vv}, we plot the finite part of the matrix element
$\mathcal{M}_{n}^{(2)}(m=0)$, defined in~\eqref{AMPvv}, computed in
the {\sc fdh} scheme, using the same convention to define the finite
part as in \cite{Ulrich:2020frs}.  Rather than showing the dependence
in the whole $s$-$t$ plane, we focus on the kinematic ranges we will
use in Section~\ref{sec:results}, i.e.~we fix $s=174\,684\,{\rm
  MeV}^2$ and consider the range $t \in [-153\,069, -1\,021]\,{\rm
  MeV}^2$.  This allows us also to demonstrate the split into the
different contributions (cf.~\eqref{AMPnnlo}).  Note that, since
$\mathcal{M}_n^{(2)}(m=0)$ is unphysical unless combined with real
corrections, we cannot make any statements about the relative sizes of
the different contributions. However, we can illustrate the numerical
stability of the analytic expressions, which are provided in an
ancillary file.

\subsubsection{Massification}
\label{sec:massif-th}

In this subsection we describe the only approximation we make for our
NNLO result. It concerns the $m$ dependence of the terms $\{q^3 Q^5,\,
q^4 Q^4,\, q^5 Q^3\}$ of $\mathcal{M}_n^{(2)}$ that are required for
$\D\sigma^{(\mathrm{vv})}_{e\mu}$. Starting from the corresponding
matrix element with a massless electron, $\mathcal{M}_n^{(2)}(m=0)$,
we can obtain the leading term of the small-mass expansion efficiently
using the strategy of massification~\cite{Penin:2005eh, Becher:2007cu,
  Engel:2018fsb}.  This way, we can recover all terms of
$\mathcal{M}_n^{(2)}(m)$ that are not polynomially suppressed,
i.e.~the logarithmically enhanced ones as well as the constant terms,
without the need of additional process-dependent computations.
However, the approximation neglects terms that vanish in the limit
$m\to0$.  Hence, in our result for the mixed corrections
$\D\sigma_{e\mu}^{(2)}$ we will miss terms of the form $(\alpha/\pi)^2
\, m^2/S$, potentially multiplied by a logarithm of the form
$\log(m^2/S)$.

Massification is applicable in the case where some external fermions
have small masses compared to all other scales in the process. Since
this corresponds to highly-energetic particles in the external states,
soft-collinear effective theory
(SCET)~\cite{Bauer:2000yr,Bauer:2001yt,Beneke:2002ph} can be used for
a systematic expansion of scattering amplitudes. As a consequence of
the decoupling transformation~\cite{Bauer:2000yr}, collinear and soft
degrees of freedom factorise at leading power and we have
\begin{align}
    \mathcal{A}_n(m)
    =\Big( \prod_j \sqrt{Z} \Big)
    \times \mathcal{S} \times \mathcal{A}_n(0)+\mathcal{O}(m)\, .
\end{align}
Each energetic external particle defines a collinear sector in SCET
and thus contributes one power of the massification constant
$\sqrt{Z}$.  This process-independent factor does not depend on any
hard scale and apart from the trivial factorised $m$ dependence is a
constant. The soft function $\mathcal{S}$, on the other hand, is not
universal and does depend on the hard scales of the process. One can
show that in QED it only receives contributions from closed fermion
loops~\cite{Becher:2007cu}. As previously mentioned, these
contributions can be calculated at NNLO with exact mass dependence
using the semi-numerical hyperspherical method. This has the added
advantage of rendering massification completely process independent.

While it is in principle possible to study these structures directly
in SCET, it is easier to instead perform a matching calculation.
In~\cite{Engel:2018fsb}, we have explicitly calculated the leading
bottom-mass effects for the process $t\to Wb$ using the method of
regions~\cite{Beneke:1997zp} and were able to write the resulting
amplitude as
\begin{align}
    \mathcal{A}_n(t\to Wb, m) = \sqrt{Z} \times \mathcal{S} \times  \mathcal{A}_n(t\to Wb, 0) + \mathcal{O}(m)\,.
\end{align}
Through trivial modifications of the colour factors, this result can
be converted to the QED case $\mu\to\nu\bar\nu e$.  The exact form of
the massification constant up to NNLO, i.e. $Z=1+Z^{(1)} + Z^{(2)}$,
can be found in~\cite{Engel:2018fsb}. For photonic corrections this
agrees with the expression given in~\cite{Becher:2007cu}.

We can now turn to our process and re-use $Z$, adding a factor of
$\sqrt{Z}$ to the amplitude for each light external leg. As mentioned
before, there is no soft contribution if closed fermion loops are
excluded. This means that for photonic corrections to muon-electron
scattering we have to write
\begin{align} \label{eq:massify}
    \mathcal{A}_n(e\mu\to e\mu, m) &= \sqrt{Z}^{\,2} \times \mathcal{A}_n(e\mu\to e\mu, 0) + \mathcal{O}(m)\,.
\end{align}
Note that $Z$ only contains contributions due to the electron charge
$q$ and is of course independent of the muon charge $Q$.  As for
\eqref{eq:polesM}, the relation \eqref{eq:massify} is to be expanded
in $\alpha$. Since we are only using massification for the mixed
contributions, it is actually sufficient to use the one-loop value of
$Z$ and we find
\begin{align}
  q^4 Q^2\, \A422(m) + q^3 Q^3\,  \A332(m) + q^2 Q^4\, \A242(m) &
  \nonumber \\
  = q^4 Q^2\, \A422(0) + q^3 Q^3\,  \A332(0) + q^2 Q^4\, \A242(0) & +
  Z^{(1)} \bigg( q^2 Q^2\, \A221(0) + q Q^3\,  \A131(0) \bigg)\,.
\end{align}
Massification was previously used~\cite{Penin:2005eh, Becher:2007cu,
  Banerjee:2021mty, Banerjee:2021qvi} to calculate mass effects in
Bhabha and M\o{}ller scattering. It was also verified in the case of
the muon decay that the massified result at NNLO gives a very good
approximation to the result with exact $m$
dependence~\cite{Engel:2019nfw}.

\subsection{Computing}

All the contributions described in the previous paragraphs are
implemented in \mcmule{}~\cite{mcmule}, a framework for (IR-safe)
fully differential higher-order QED calculations. The code is publicly
available at
\begin{center}
   \url{https://gitlab.com/mule-tools/mcmule}
\end{center}

A general presentation of \mcmule{} and of the methods employed
therein is given in \cite{Banerjee:2020rww, Ulrich:2020frs,
  Engel:2022kde}. In particular, Section~5 of \cite{Banerjee:2020rww}
deals with the implementation of the electronic corrections at
NNLO. Here the description is focused on the peculiarities of the new
muon-electron scattering implementation, which completes the previous
one.

The customary split of the radiative corrections into fermionic and
photonic contributions is also reflected by the structure of the
code. The user is allowed to choose which contributions, or pieces in
the \mcmule{} notation, are to be computed. Each of the two classes is
further split in terms of electronic, muonic and mixed corrections.

The non-radiative pieces are relatively straightforward to integrate
over their simple phase space as there are no numerical issues.
However, there is one subtlety related to the double-virtual mixed
photonic piece: the matrix element is expressed in terms of several
thousand GPLs.  While these can be efficiently evaluated using the
latest version ({\tt v0.2.0b}) of {\tt handyG}~\cite{Naterop:2019xaf},
evaluation is still limited to roughly $1\,{\rm s}/{\rm event}$.
However, due to the simplicity of the actual phase-space integration
that only requires $\sim 2\times 10^5$ points, it is still possible to
obtain very good results in a reasonable amount of time.  A more
aggressive and efficient caching system in {\tt handyG} might improve
this further.

The radiative pieces have comparatively simpler matrix elements but a
more complicated phase space that tends to lead to numerical
instabilities.  The main issue here is the real-virtual matrix element
which is evaluated using {\sc OpenLoops} and NTS stabilisation (see
Section~\ref{sec:OLrad}).  For points in the bulk of the phase space,
evaluation takes roughly $3\,{\rm ms}/{\rm event}$.  However, the
complexity of the phase space necessitates roughly $1.5\times 10^8$
points meaning that the total integration time is on par with the
non-radiative pieces.

The whole set of results can be found in the relevant directory of the
\mcmule{} user library,
\begin{center}
    \url{https://mule-tools.gitlab.io/user-library/}
\end{center}
along with user, menu and configuration files, and the Python code
that generates the plots in the paper~\cite{McMule:data}.  The
production runs employed version {\tt v0.3.1} (for $\mu^-e\to\mu^-e$)
and {\tt v0.4.2} (for $\mu^+e\to\mu^+e$) of the \mcmule{} public
release, which was encapsulated in a {\tt docker} environment in order
to ensure complete reproducibility of the results.  In order to obtain
a relative accuracy better than at least $10^{-4}$ ($10^{-3}$) for the
NLO (NNLO) coefficient, we ran \mcmule{} for a total of 2.5 CPU-years
per configuration on Intel Xeon Gold 6152 CPUs.  In total, the results
presented in this paper, including the checks in
Section~\ref{sec:check}, correspond to a runtime of roughly 21
CPU-years.\footnote{Using~\cite{Lannelongue:2021co2} we estimate that
  this corresponds to an energy consumption of 2.77 MWh and a carbon
  footprint of 32 kgCO$_2$e.}

%% file: 03_results.tex

\section{Results}
\label{sec:results}

This section presents some results for muon-electron scattering at
NNLO, with the characteristics of the MUonE experiment in mind.  The
kinematics of the process is defined by the momenta
in~\eqref{eq:kinematics} together with the electron and the muon mass,
$m$ and $M$. The Mandelstam invariants are introduced as
$t_e=(p_1-p_3)^2$ and $t_\mu=(p_2-p_4)^2$, along with the outgoing
electron and muon energy, $E_e$ and $E_\mu$, and the electron and muon
scattering angle with respect to the beam axis, $\theta_e$ and
$\theta_\mu$. If the process is elastic $t_e = t_\mu \equiv t$.

The muon beam energy is set to $E=160$\,GeV, consistent with the M2
beam line at CERN North Area~\cite{Spedicato:2022qtw}.\footnote{This
  is different than the value used in previous
  works~\cite{Banerjee:2020rww,CarloniCalame:2020yoz} where
  $E=150$\,GeV.} This corresponds to a centre-of-mass energy of
$\sqrt{s} \approx 420\,\mathrm{MeV}$. A cut is imposed on the energy
of the outgoing electron, $E_e > 1$\,GeV, which is equivalent to a cut
on the minimal value of $|t|$, in order to cure the singular behaviour
of $\D\sigma/\D t \sim t^{-2}$. A cut on $\theta_\mu$ can be used to
remove most of the background. Hence, for some of our results we also
require $\theta_\mu > 0.3$\,mrad.

In order to achieve a well-defined extraction of the HVP, a possible
way to proceed is to discriminate elastic scattering events from the
otherwise kinematically allowed radiative events and processes in the
background. This can be obtained in terms of the elasticity constraint
that relates muon and electron scattering angles in the absence of
photons,
\begin{equation}
    \tan\theta^{\rm el}_\mu = \frac{2\tan\theta_e}{(1+\gamma^2
      \tan^2\theta_e) (1+g^*_\mu)-2} \,,
\end{equation}
where
\begin{equation}
    \gamma = \frac{E+m}{\sqrt{s}}\,, \qquad g^*_\mu = \frac{E m+M^2}{E
      m+m^2}\,.
\end{equation}
Applying an elasticity
cut, such as
\begin{equation} \label{ec:elcut}
    0.9 < \frac{\theta_\mu}{\theta^{\rm el}_\mu} < 1.1 \,,
\end{equation}
would then allow to reconstruct the HVP momentum flow on an
event-by-event basis. Furthermore, the cut is expected to flatten out
the radiative corrections at the differential level due to an evenly
distributed soft enhancement. Since MUonE plans to exploit the
presence of a normalisation and a signal region, this behaviour could
turn out to be advantageous. The effect of the elasticity cut
\eqref{ec:elcut} is very similar to the acoplanarity cut introduced
in~\cite{Alacevich:2018vez}.

However, such a kinematical constraint is not ideal from the
experimental perspective. It would cut off many events, yielding
issues in terms of statistics, and would also complicate the estimate
of systematic uncertainties, as it would lead to a complex practical
implementation. At present, the alternative proposed by the experiment
is to employ a template fit to extract the HVP, as discussed
in~\cite{Abbiendi:2022oks}. Nonetheless, a study with the elasticity
cut is still of theoretical interest.

In the following, results are presented for different scenarios,
defined in terms of the kinematical cuts discussed above and
summarised in Table \ref{tab:scenarios}.
\begin{table}[h]
    \centering
    \renewcommand{\arraystretch}{1.3}
    \begin{tabular}{l|c|c|c}
                  & $E_e > 1$ GeV & $\theta_\mu > 0.3$ mrad & $0.9 < \theta_\mu/\theta^{\rm el}_\mu < 1.1$  \\
        \hline
        {\tt S1}  & \checkmark & \checkmark  &  \\
        \hline
        {\tt S2}  & \checkmark & \checkmark  & \checkmark \\
        \hline
        {\tt S1}' & \checkmark &             &  \\
        \hline
        {\tt S2}' & \checkmark &             & \checkmark
    \end{tabular}
    \caption{Kinematical scenarios analysed in the \mcmule{}
      prediction.}
    \label{tab:scenarios}
\end{table}
All the results use the input parameters~\cite{Workman:2022ynf}
\begin{align*}
    \alpha &= 1/137.035999084      \,,&  m &=  0.510998950 \,\,{\MeV} \,, \\
    M  & =  105.658375 \,\,{\MeV} \,,&  m_\tau &= 1776.86     \,\,{\MeV} \,.
\end{align*}
The Fortran library {\tt alphaQED}~\cite{alphaqed}, in particular the
most recent version {\tt alphaQEDc19}, is employed for the evaluation
of the HVP.

The order-by-order contributions, $\sigma^{(i)}$, to the N$^{k}$LO
integrated cross section, $\sigma_k = \sum_{i=0}^k \sigma^{(i)}$, are
shown in Tables~\ref{tab:tot_xsec} and~\ref{tab:tot_xsecp}, divided
according to~\eqref{SGI}. The NLO hadronic piece, $\sigma^{(1)}_{\rm
  had}$, corresponds to the signal of the MUonE experiment. In
addition, the tables show the $K$ factor corresponding to each
contribution, defined as
\begin{align}
    K^{(i)} = 1+\delta K^{(i)} = \frac{\sigma_i}{\sigma_{i-1}} \,.
\end{align}

When applicable, the tables present the different cross sections for
both negative and positive muons. The mixed photonic NLO correction
with positive muons can be derived from the one with negative muons by
flipping the sign of the latter. The mixed photonic NNLO correction
could be further disentangled into three classes, labelled by the
leptonic charges relative to the LO cross section, i.e.~$\{q^3\,Q,\,
q^2\,Q^2,\, q\,Q^3\}$, so that the contribution with positive muons
can be derived from the one with negative muons by flipping the sign
of the classes with odd powers of $Q$. Once all mixed contributions
are added up as displayed in the tables, this symmetry is not manifest
anymore.

\begin{table}[ht]
    \centering
    \renewcommand{\arraystretch}{1.3}
    \begin{tabular}{l|rr|rr}
        &\multicolumn{2}{ c}{$\sigma / \upmu{\rm b}$}
        &\multicolumn{2}{|c}{$\delta K^{(i)} / \%$}               \\
               & {\tt S1\phantom{aaaa}} & {\tt S2\phantom{aaaa}}
               & {\tt S1\phantom{aaaa}} & {\tt S2\phantom{aaaa}} \\
        \hline
        $\szt  $&$ 106.44356\pp $&$ 106.44356\pp $&               &               \\
        \hline
        $\soe  $&$  -0.61211(3) $&$  -4.66042(3) $&$ -0.57506(3) $&$ -4.37830(3) $\\
        \multirow{2}{*}{
        $\sox$} &$  -0.21404\pp $&$  -0.16017\pp $&$ -0.20108\pp $&$ -0.15047\pp $\\
                &$   0.21404\pp $&$   0.16017\pp $&$  0.20108\pp $&$  0.15047\pp $\\
        $\som  $&$  -0.02843\pp $&$  -0.16134\pp $&$ -0.02671\pp $&$ -0.15157\pp $\\
        $\sol  $&$   1.38575\pp $&$   1.38575\pp $&$  1.30186\pp $&$  1.30186\pp $\\
        $\soh  $&$   0.01565\pp $&$   0.01565\pp $&$  0.01471\pp $&$  0.01471\pp $\\
        \hline
        \multirow{2}{*}{
        $\sot$} &$ 106.99038(3) $&$ 102.86304(3) $&$  0.51372(3) $&$ -3.36377(3) $\\
                &$ 107.41847(3) $&$ 103.18338(3) $&$  0.91589(3) $&$ -3.06283(3) $\\
        \hline
        $\ste  $&$   0.00090\pp $&$   0.06595\pp $&$  0.00084\pp $&$  0.06411\pp $\\
        \multirow{2}{*}{
        $\stx$} &$   0.00095\pp $&$   0.01926\pp $&$  0.00089\pp $&$  0.01872\pp $\\
                &$   0.00329\pp $&$   0.00479\pp $&$  0.00307\pp $&$  0.00464\pp $\\
        $\stm  $&$  -0.00005\pp $&$   0.00002\pp $&$ -0.00005\pp $&$  0.00002\pp $\\
        \multirow{2}{*}{
        $\stl$} &$  -0.01195\pp $&$  -0.06568\pp $&$ -0.01117\pp $&$ -0.06385\pp $\\
                &$  -0.00424\pp $&$  -0.05959\pp $&$ -0.00395\pp $&$ -0.05775\pp $\\
        \multirow{2}{*}{
        $\sth$} &$  -0.00045\pp $&$  -0.00104\pp $&$ -0.00042\pp $&$ -0.00101\pp $\\
                &$  -0.00004\pp $&$  -0.00068\pp $&$ -0.00004\pp $&$ -0.00066\pp $\\
        \hline
        \multirow{2}{*}{
        $\stt$} &$ 106.97977(3) $&$ 102.88154(3) $&$ -0.00992(4) $&$  0.01799(4) $\\
                &$ 107.41832(3) $&$ 103.19386(3) $&$ -0.00013(4) $&$  0.01016(4) $
    \end{tabular}
    \caption{Integrated cross sections for {\tt S1} and {\tt S2} at
      LO, NLO, and NNLO. The results are split into photonic,
      i.e.~electronic, mixed and muonic, and fermionic, i.e.~leptonic
      and hadronic, corrections. All three leptons are included in the
      leptonic contributions. When applicable, the different
      contributions with negative and positive muons are shown. Where
      no error is given, all digits are significant compared to the
      precision of the numerical integration.}
    \label{tab:tot_xsec}
\end{table}
\begin{table}[ht]
    \centering
    \renewcommand{\arraystretch}{1.3}
    \begin{tabular}{l|rr|rr}
        &\multicolumn{2}{ c}{$\sigma / \upmu{\rm b}$}
        &\multicolumn{2}{|c}{$\delta K^{(i)} / \%$}                \\
                & {\tt S1'\phantom{aaa}} & {\tt S2'\phantom{aaa}}
                & {\tt S1'\phantom{aaa}} & {\tt S2'\phantom{aaa}} \\
        \hline
        $\szt  $&$ 245.59625(1)   $&$ 245.59625(1) $&               &             \\
        \hline
        $\soe  $&$  10.8488(2)\po $&$ -11.18183(7) $&$  4.41735(7) $&$ -4.55293(3) $\\
        \multirow{2}{*}{
        $\sox$} &$  -0.35021\pp   $&$  -0.28598\pp $&$ -0.14260\pp $&$ -0.11644\pp $\\
                &$   0.35021\pp   $&$   0.28598\pp $&$  0.14260\pp $&$  0.11644\pp $\\
        $\som  $&$  -0.06668\pp   $&$  -0.22413\pp $&$ -0.02715\pp $&$ -0.09126\pp $\\
        $\sol  $&$   2.88969\pp   $&$   2.88969\pp $&$  1.17660\pp $&$  1.17660\pp $\\
        $\soh  $&$   0.01943\pp   $&$   0.01943\pp $&$  0.00791\pp $&$  0.00791\pp $\\
        \hline
        \multirow{2}{*}{
        $\sot$} &$ 258.9373(2)\po $&$ 236.81343(7) $&$  5.43211(7) $&$ -3.57612(3) $\\
                &$ 259.6377(2)\po $&$ 237.38539(7) $&$  5.71731(7) $&$ -3.34324(3) $\\
        \hline
        $\ste  $&$   0.02713(4)   $&$   0.17491(1) $&$  0.01048(2) $&$  0.07386\pp $\\
        \multirow{2}{*}{
        $\stx$} &$  -0.02526\pp   $&$   0.02852\pp $&$ -0.00975\pp $&$  0.01204\pp $\\
                &$   0.03165\pp   $&$   0.00439\pp $&$  0.01219\pp $&$  0.01852\pp $\\
        $\stm  $&$  -0.00010\pp   $&$  -0.00004\pp $&$ -0.00004\pp $&$ -0.00002\pp $\\
        \multirow{2}{*}{
        $\stl$} &$   0.05911\pp   $&$  -0.13445\pp $&$  0.02283\pp $&$ -0.05678\pp $\\
                &$   0.07161\pp   $&$  -0.12380\pp $&$  0.02758\pp $&$ -0.05215\pp $\\
        \multirow{2}{*}{
        $\sth$} &$  -0.00049\pp   $&$  -0.00128\pp $&$ -0.00019\pp $&$ -0.00054\pp $\\
                &$   0.00000\pp   $&$  -0.00083\pp $&$  0.00000\pp $&$ -0.00035\pp $\\
        \hline
        \multirow{2}{*}{
        $\stt$} &$ 258.9977(2)\po  $&$ 236.88109(7) $&$  0.02332(9) $&$  0.02857(4) $\\
                &$ 259.7680(2)\po  $&$ 237.44002(7) $&$  0.05018(9) $&$  0.02301(4) $
    \end{tabular}
    \caption{Integrated cross sections for {\tt S1}' and {\tt S2}' at
      LO, NLO, and NNLO. The results are split into photonic,
      i.e.~electronic, mixed and muonic, and fermionic, i.e.~leptonic
      and hadronic, corrections. All three leptons are included in the
      leptonic contributions. When applicable, the different
      contributions with negative and positive muons are shown. Where
      no error is given, all digits are significant compared to the
      precision of the numerical integration.}
    \label{tab:tot_xsecp}
\end{table}

According to Table~\ref{tab:tot_xsec}, NLO and NNLO corrections amount
to around 1\% and 0.01\% (or less) for {\tt S1}, and to around 3\% and
0.01\% for {\tt S2}.  The elasticity constraint cuts off hard
radiation and the consequent soft enhancement introduces large
logarithms that result in larger $K$ ratios in the latter scenario,
particularly in the case of electronic corrections.  For {\tt S1}
there is no apparent hierarchy among photonic corrections, while
leptonic contributions, especially those due to electronic VP
insertions, dominate NLO and NNLO corrections. For {\tt S2} a
hierarchy between electronic and mixed photonic contributions, both at
NLO and at NNLO, is more evident.

Comparing Table~\ref{tab:tot_xsecp} and Table~\ref{tab:tot_xsec}
reveals that the cut on $\theta_\mu$ has a large impact on NLO
corrections, similar to the elasticity cut. However, the $\theta_\mu$
cut has a much more pronounced effect on the total cross section, as
it also affects the Born term.

\begin{figure}
    \centering
    \includegraphics[width=0.9\textwidth]{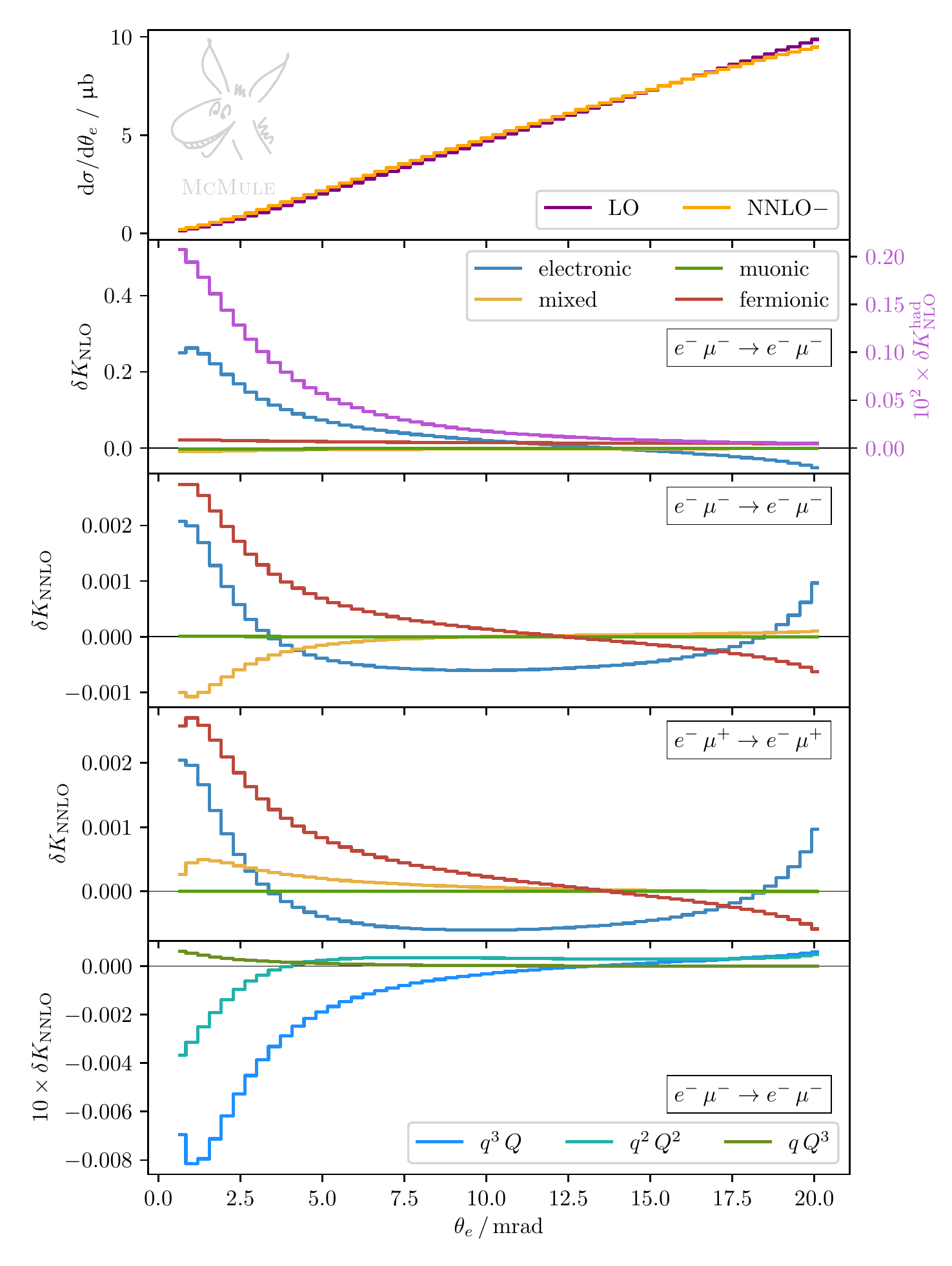}
    \caption{From top to bottom: (i) differential cross section
      w.r.t.~$\theta_e$ for {\tt S1} at LO (violet), and NNLO (orange)
      for negative muons; (ii) NLO $K$ factor for negative muons
      (positive muons have a sign flip for the mixed photonic
      correction); (iii) NNLO $K$ factor for negative muons; (iv) NNLO
      $K$ factor for positive muons; (v) NNLO $K$ factor for
      disentangled mixed photonic corrections, for negative muons. In
      panels~(ii)--(iv) the correction is split into photonic,
      i.e.~electronic, mixed and muonic, and fermionic, including
      leptonic and hadronic. The hadronic correction at NLO
      corresponds to the signal of the experiment and is shown
      separately in purple in panel~(ii).}
    \label{fig:dsdthe0}
\end{figure}

\begin{figure}
    \centering
    \includegraphics[width=0.9\textwidth]{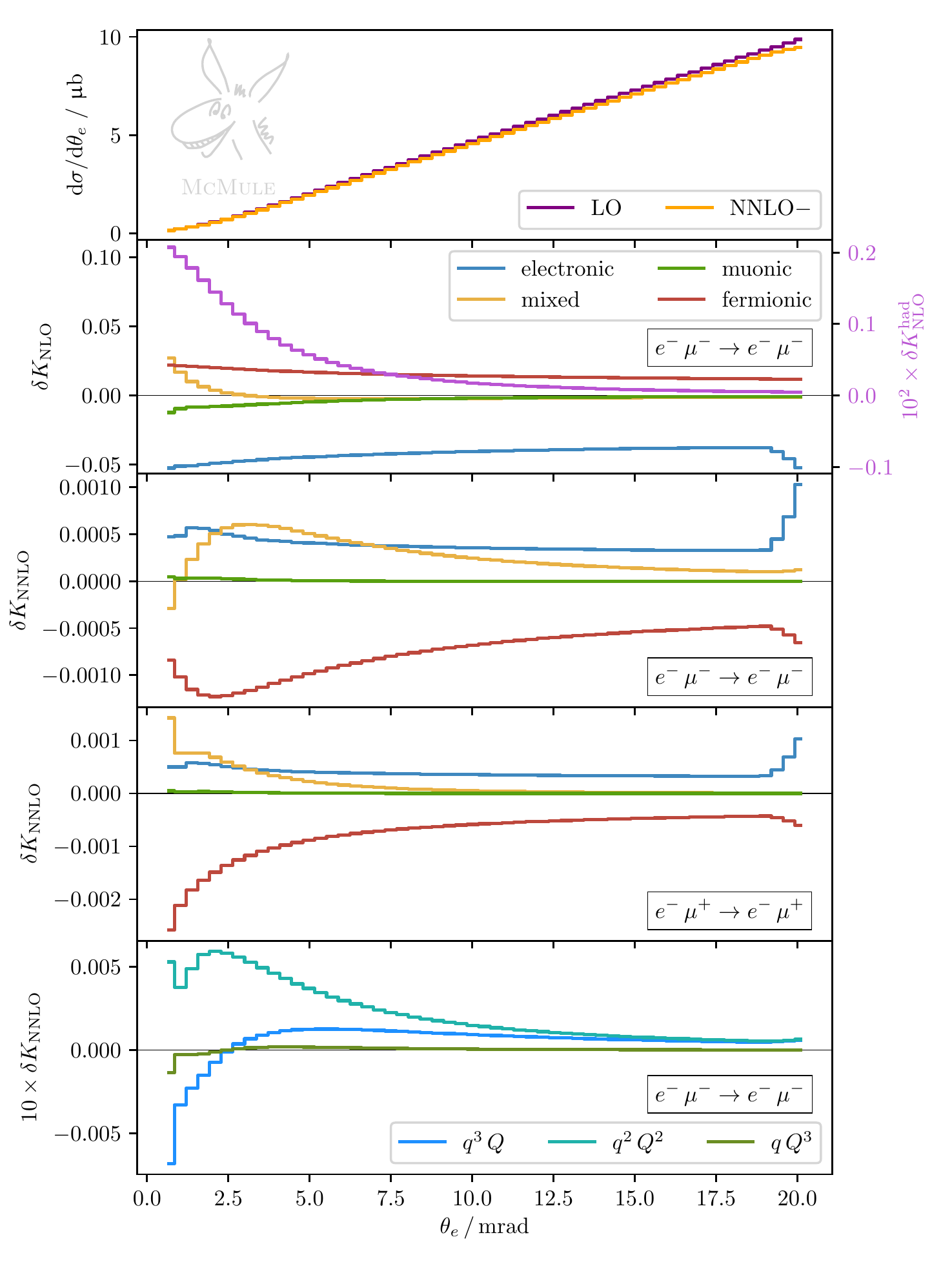}
    \caption{From top to bottom: (i) differential cross section
      w.r.t.~$\theta_e$ for {\tt S2} at LO (violet), and NNLO (orange)
      for negative muons; (ii) NLO $K$ factor for negative muons
      (positive muons have a sign flip for the mixed photonic
      correction); (iii) NNLO $K$ factor for negative muons; (iv) NNLO
      $K$ factor for positive muons; (v) NNLO $K$ factor for
      disentangled mixed photonic corrections, for negative muons. In
      panels~(ii)--(iv) the correction is split into photonic,
      i.e.~electronic, mixed and muonic, and fermionic, including
      leptonic and hadronic. The hadronic correction at NLO
      corresponds to the signal of the experiment and is shown
      separately in purple in panel~(ii).}
    \label{fig:dsdthe1}
\end{figure}

\begin{figure}
    \centering
    \includegraphics[width=0.9\textwidth]{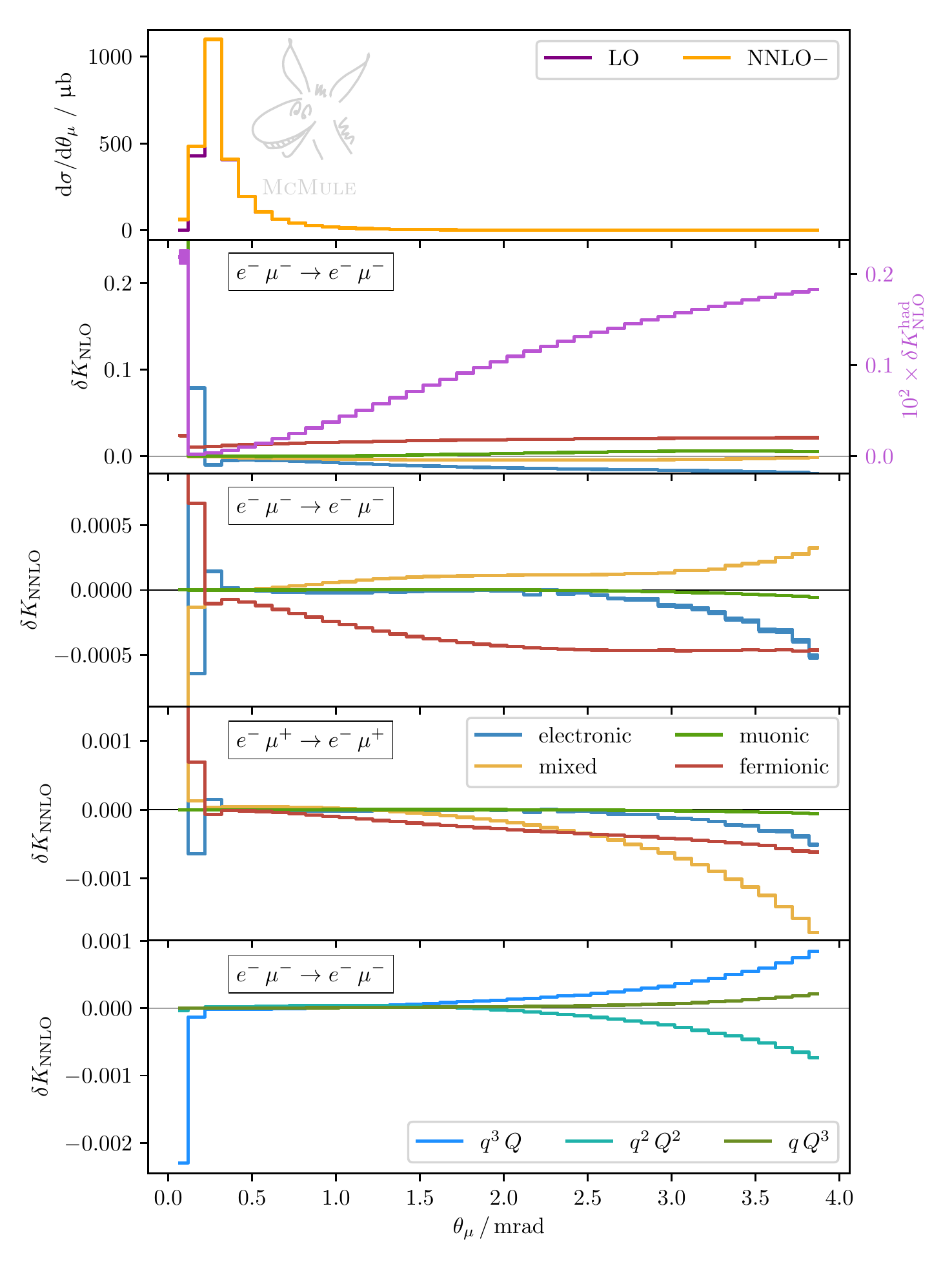}
    \caption{From top to bottom: (i) differential cross section
      w.r.t.~$\theta_\mu$ for {\tt S1}' at LO (violet), and NNLO
      (orange) for negative muons; (ii) NLO $K$ factor for negative
      muons (positive muons have a sign flip for the mixed photonic
      correction); (iii) NNLO $K$ factor for negative muons; (iv) NNLO
      $K$ factor for positive muons; (v) NNLO $K$ factor for
      disentangled mixed photonic corrections, for negative muons. In
      panels~(ii)--(iv) the correction is split into photonic,
      i.e.~electronic, mixed and muonic, and fermionic, including
      leptonic and hadronic. The hadronic correction at NLO
      corresponds to the signal of the experiment and is shown
      separately in purple in panel~(ii).}
    \label{fig:dsdthm0}
\end{figure}

In addition to integrated cross sections, differential distributions
can provide a more reliable estimate with respect to MUonE's 10 ppm
target, as higher-order corrections can be much larger at that level.
As a Monte Carlo integrator, \mcmule{} allows for the calculation of
any number of IR-safe differential observables in the same run.
Here,\footnote{Results for other observables can be found in the
  \mcmule{} \href{https://mule-tools.gitlab.io/user-library/}{User
    Library}.}  Figures~\ref{fig:dsdthe0}--\ref{fig:dsdthm0} display
differential results that are of interest to the MUonE experiment, in
particular distributions with respect to $\theta_e$ and
$\theta_\mu$. The differential cross sections at LO and NNLO are shown
in the upper panels. Furthermore, the lower panels show the
differential $K$ factor
\begin{align}
    \delta K^{(i)} =
    \frac{\D\sigma^{(i)}/\D x}{\D\sigma_{i-1}/\D x}
    \,,
\end{align}
with $x\in\{\theta_e,\theta_\mu\}$. From top to bottom, the second
panel displays the NLO $K$ factor for negative muons, the third and
fourth panel the NNLO $K$ factor for negative and positive muons, the
fifth panel again the NNLO $K$ factor for negative muons restricted to
mixed photonic contributions, further disentangled in terms of the
three gauge-invariant subsets, labelled by their leptonic charge.

In the case of the $\theta_e$ distribution, (N)NLO corrections can be
larger at differential level, for example up to 20\% (0.2\%), as shown
in Figure~\ref{fig:dsdthe0}. These large corrections occur for small
electron scattering angles, or equivalently for large electron
energies, where photon emission is forced to be soft. Furthermore, the
effect of the elasticity cut is clearly visible when comparing
Figure~\ref{fig:dsdthe0} and Figure~\ref{fig:dsdthe1}. As expected
from an evenly-distributed soft enhancement, the $K$ factor is
significantly flattened.

Among photonic corrections, a hierarchy is expected from the
appearance of collinear pseudo-singularities. At the cross section
level, this introduces logarithms of the form $\log(m_i^2/S)$, where
$m_i \in \{m,\,M\}$ and $S$ is the energy scale of the process. As a
consequence, electronic corrections are expected to be dominant
compared to mixed corrections, and even more compared to muonic
corrections.

This phenomenon can be observed at NLO, where the electronic
contribution is the largest, and particularly for {\tt S1}, where the
additional soft enhancement at the endpoints of the distributions
seems to affect the electronic correction even more.

The collinear hierarchy is less pronounced at NNLO, apparently because
of further enhancements of soft origin. In fact, electronic and mixed
corrections are similar in magnitude for small $\theta_e$, where the
soft enhancement is prevalent. Nonetheless, the collinear hierarchy is
partially restored in the bulk of the distribution.  The interplay
between collinear and soft enhancements is even clearer from the
lowest panel in Figure~\ref{fig:dsdthe0}, where the soft enhancement
of the mostly-electronic mixed correction~($q^3 Q$) turns out to
account for the similar magnitude of electronic and mixed corrections
for small $\theta_e$.  Similarly, NNLO results for {\tt S2} in
Figure~\ref{fig:dsdthe1} are further enhanced by soft logarithms
because of the additional elasticity constraint. As a consequence, the
collinear hierarchy is less visible in the bulk of the distribution.

In the case of the $\theta_\mu$ distribution in {\tt S1'}, displayed
in Figure~\ref{fig:dsdthm0}, a strong enhancement for small values of
$\theta_\mu$ is visible for all NLO and NNLO $K$ factors, except for
the NLO fermionic. This explains why MUonE enforces a kinematical cut
for small muon scattering angles, in order not to lose the majority of
these events. On the other hand, fermionic corrections at NLO are not
enhanced~(note the different scale for the NLO hadronic), since NLO
fermionic corrections correspond to elastic events, which are not
affected by the cut at $\theta_\mu = 0.3$ mrad.

Finally, Figure~\ref{fig:dsdthe0} and Figure~\ref{fig:dsdthm0} also
show that the MUonE signal, i.e.~the hadronic NLO corrections, are
$\mathcal{O}(10^{-3})$ in the regions of small $\theta_e$ and large
$\theta_\mu$.

However, the magnitude of NNLO corrections at differential level,
around $10^{-3}$, is still too large compared to MUonE's precision
goal of 10 ppm. Higher-order predictions beyond NNLO can certainly
help in that direction. Furthermore, it is clearly mandatory to make
an effort towards a more reliable description of the region where
radiation leads to an enhancement through large logarithms.

%% file: 04_checks.tex

\section{Checks and validation}
\label{sec:check}

Since this paper presents the first full prediction for muon-electron
scattering at NNLO in QED, various studies were conducted in order to
validate the results. Checks were performed both externally
(Section~\ref{sec:external}), against independent though partial
results, and internally (Section~\ref{sec:internal}), in order to test
the consistency and the validity of the methods.  In particular, both
photonic and fermionic NNLO corrections could be completely tested
against independent calculations, except for the contribution of the
two-loop four-point topologies to the mixed photonic correction.

\subsection{External checks}
\label{sec:external}

The electronic and muonic contributions to the photonic NNLO
correction were also calculated in~\cite{CarloniCalame:2020yoz} with
{\sc Mesmer}. Since a slicing scheme is employed therein to handle IR
divergences, along with a photon-mass regulator, a comparison with
\mcmule{} represents a completely independent check. A positive
outcome was reported in~\cite{Banerjee:2020rww}. As mentioned in
Section~\ref{sec:OLrad}, we have improved upon this calculation by
using {\sc OpenLoops} combined with NTS stabilisation to evaluate the
real-virtual contribution. This yields a better convergence behaviour
of the Monte Carlo integration. Perfect agreement with the {\sc
  Mesmer} results is also found in this case.

For the mixed photonic NNLO result, a new comparison was conducted
with the {\sc Mesmer} collaboration. Since the calculation
in~\cite{CarloniCalame:2020yoz} is complete up to the mixed two-loop
contribution, it is possible to compare the mixed NLO correction to
$\mu e\to \mu e\gamma$, which is physical and corresponds to the
double-real and real-virtual contributions to muon-electron
scattering. In order to check the numerical stability of the
real-virtual implementation, small photon energy cuts of
$\{10^{-6},10^{-5},10^{-4}\}\times \sqrt{s}/2$ were used. Perfect
agreement was found between the two codes for the total cross section
as well as for differential distributions.

A verification of the complete NNLO calculation and in particular of
the genuine two-loop four-point topologies with {\sc Mesmer} is
currently not possible. In~\cite{CarloniCalame:2020yoz} these
contributions were approximated with a YFS-inspired approach. Since
this only yields a rough estimate of the exact result, a meaningful
comparison is not viable. Instead, we have made an independent test of
the massified two-loop matrix element. To this end, we start from a
completely independent massless calculation of the matrix element for
Bhabha scattering~\cite{Bern:2000ie}.  We restrict the result to the
$t$-channel~\cite{Banerjee:2021mty} contribution and massify both
fermion lines. As can be expected, this result approaches the
muon-electron two-loop matrix element in the high-energy limit $S\gg
M,m$. This represents a strong check on the correctness of the
massless two-loop matrix element as well as the consistency of the
massification procedure.

Finally, the fermionic contribution to the NNLO correction was also
computed in~\cite{Fael:2019nsf} through a dispersive approach, using
$e^+ e^-$ annihilation data for the hadronic part. On the other hand,
the \mcmule{} implementation employs the hyperspherical results
from~\cite{Fael:2018dmz}. The positive outcome of the comparison
between the two sets of results, for both leptonic and hadronic
distributions, constitutes an additional strong validation.

\subsection{Internal checks}
\label{sec:internal}

Further checks were performed internally to test the new ingredients
of the present calculation and are discussed with more details in the
following subsections. In particular, Section~\ref{sec:impl} presents
a validation for the Fortran implementation in \mcmule{},
Section~\ref{sec:massif} analyses the effect of massification, and
Section~\ref{sec:nts} studies the impact of NTS stabilisation applied
to the calculation of real-virtual contributions. The results of all
of these checks are summarised in Figure~\ref{fig:val-tot}.

\subsubsection{Implementation} \label{sec:impl}

In addition to the external checks, a validation for the Fortran
implementation is provided through the subtraction method employed by
\mcmule{}. As given in~\eqref{FKSnnlo}, FKS$^\ell$ introduces the
unphysical parameter $\xi_c$, on which the three pieces contributing
to the photonic NNLO correction (double-virtual, real-virtual and
double-real) depend. The cancellation of the $\xi_c$ dependence, when
the three pieces are summed together, is a strong check in terms of
implementation consistency and numerical stability, even more when
performed at the level of differential distributions.

\begin{figure}
    \centering
    \includegraphics[width=0.9\textwidth]{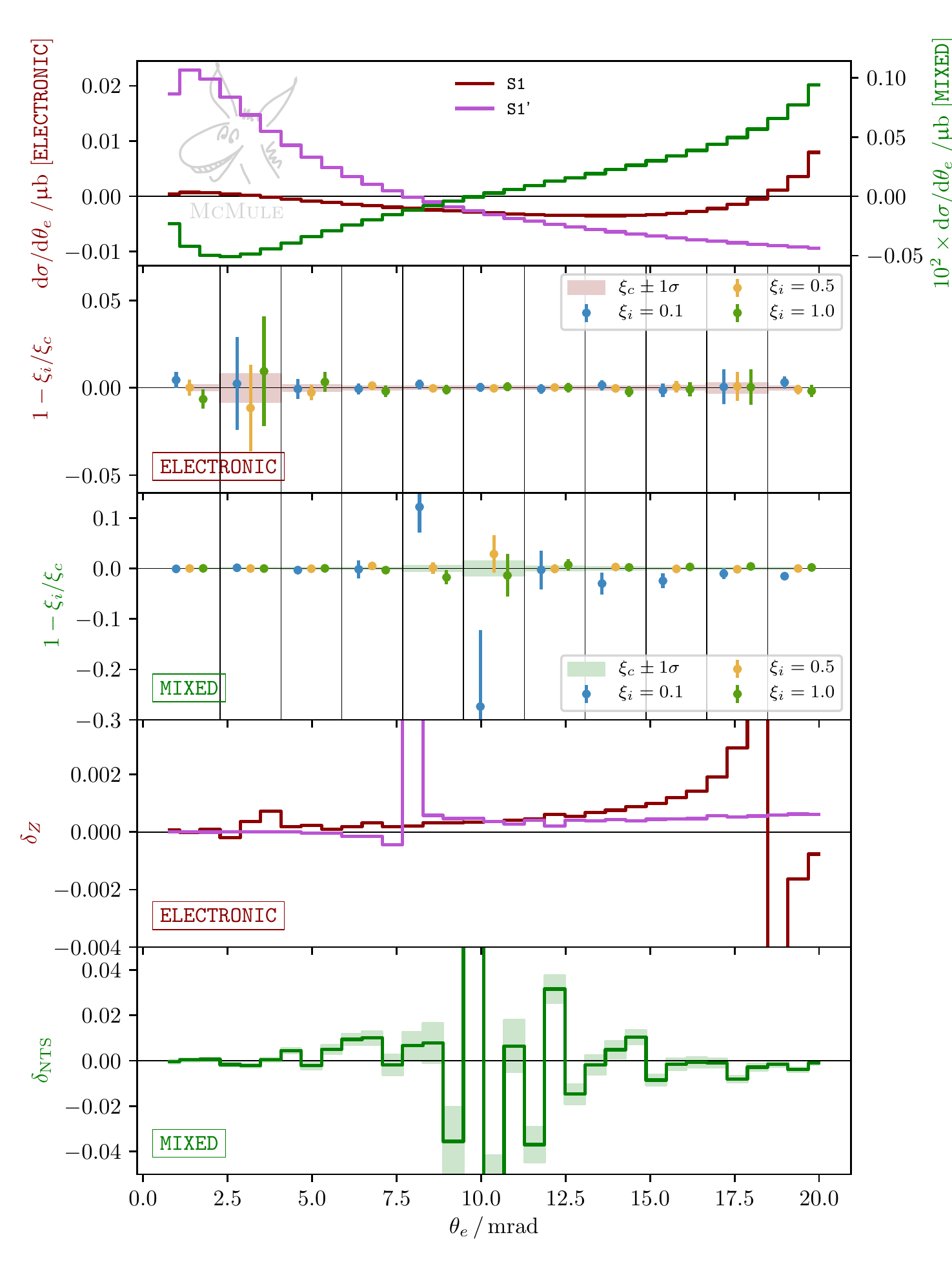}
    \caption{Validation plots for electronic and mixed photonic
      corrections, as differential distributions w.r.t.~the electron
      scattering angle. From top to bottom~(see the corresponding
      subsections in Section~\ref{sec:check} for further details): (i)
      differential NNLO correction w.r.t.~$\theta_e$ -- electronic in
      {\tt S1} in red, electronic in {\tt S1}' in purple, mixed in
      {\tt S1} in green; (ii) $\xi_c$-(in)dependence study for NNLO
      electronic corrections in {\tt S1}; (iii) $\xi_c$-(in)dependence
      study for NNLO mixed corrections in {\tt S1}; (iv) massification
      study for NNLO electronic corrections -- {\tt S1} in red, {\tt
        S1}' in purple; (v) NTS stabilisation study for NNLO mixed
      corrections.  See~\eqref{eq:val_z} and~\eqref{eq:val_nts} for
      details on $\delta_Z$ and $\delta_{\rm NTS}$.}
    \label{fig:val-tot}
\end{figure}

The second and third panel of Figure~\ref{fig:val-tot} show, for a
number of merged bins, the relative difference between the
differential cross section computed for each of the $\xi_c$ employed
for the calculation ($\xi_i$) and the combined differential cross
section (labelled with $\xi_c$), for the electronic and the mixed NNLO
correction in {\tt S1}, respectively.

The error bars in the plot indicate very good consistency and
stability of the implementation, on top of~(unavoidable) larger
oscillations corresponding to the zero crossings of the distributions.

\subsubsection{Massification error} \label{sec:massif}

It is essential to validate the reliability of the approximation
provided by massification for the considered observables. To do so,
one can use the electronic NNLO correction, computed with full mass
dependence, to obtain an error estimate when massification is applied
to the massless two-loop electron form factor.

We write the massified electronic NNLO corrections as
\begin{align}
\label{Pdef:mfied}
 \D\tilde\sigma_e^{(2)}(m) &= \big(1+\delta_Z \big)\, \D\sigma_e^{(2)}(m)
\end{align}
and thereby introduce the relative difference, $\delta_Z$, between the
NNLO coefficients of the full and massified electronic correction.
The second-to-last panel of Figure~\ref{fig:val-tot} shows $\delta_Z$
in two different scenarios, {\tt S1} and {\tt S1}', which differ for
the enforcement of an additional cut on the muon scattering angle,
$\theta_\mu > 0.3$ mrad $\equiv\theta_\mu^c$.  We stress that the
difference between the massified and full result for the NNLO
coefficients only affects the double-virtual contributions, which are
computed with and without massification, respectively.

The relative difference at the level of the NNLO differential cross
section is
\begin{align}
\label{eq:val_z}
  \frac{\D\tilde\sigma_2 - \D\sigma_2}{\D\sigma_2} &=
  \frac{\D\sigma^{(2)}_e}{\D\sigma_2}  \, \delta_Z
  \sim \alpha^2 \, \delta_Z \,,
\end{align}
i.e.~at most around $10^{-3} \alpha^2$ in the present case, so that we
can safely argue that massification is not providing a source of
uncertainty in the context of MUonE's precision goal of 10 ppm.  The
error bars are not visible since they are much smaller than the scale
of $\delta_Z$. In particular, they quantify the error of the ratio
between the double-virtual contributions to the massified and the full
NNLO correction. This error is at least a thousand times smaller than
the Monte Carlo error of the physical NNLO corrections, not shown in
the plot.

However, in the case of {\tt S1}, $\delta_Z$ seems to increase for
$\theta_e \sim 15$ mrad, on top of~(unavoidable) larger oscillations
due to the zero crossings of the distributions.  Since massification
is applied to the double-virtual contribution, this behaviour can be
explained with a focus on elastic events. In this respect, electron
scattering angles around 15 mrad correspond to muon scattering angles
around 0.5 mrad. As $\theta_e$ increases, the difference between the
corresponding $\theta_\mu$ and $\theta_\mu^c$ gets smaller and
smaller. The appearance of such a small quantity breaks the power
counting massification is based on and thus results in a less reliable
approximation. This observation is confirmed by the absence of a
similar trend in the case of {\tt S1}', where the same cut is not
enforced, in purple in Figure~\ref{fig:val-tot}.

\subsubsection{Impact of next-to-soft stabilisation} \label{sec:nts}

A final check validates the procedure of NTS stabilisation, which is
used for computing real-virtual contributions in phase-space regions
where the evaluation with {\sc OpenLoops} becomes numerically
unstable.

Below a certain value of the photon energy, fixed at $10^{-3}
\sqrt{s}/2$ for the present implementation, the evaluation of the
real-virtual matrix element is switched to the NTS expansion.  This is
a much simpler and thus numerically much more stable expression.
Alternatively, one can use {\sc OpenLoops} in quadruple precision to
always use the full matrix element for the entire phase space. This
has an enormous price in terms of computing time, but can be used to
validate the NTS expansion.

In analogy to \eqref{Pdef:mfied} we write the NTS-expanded mixed NNLO
correction as $(1+\delta_{\rm NTS})\,\D\sigma_{e\mu}^{(2)}$, where
$\D\sigma_{e\mu}^{(2)}$ is computed using {\sc OpenLoops} in quadruple
precision for the full phase space.  The last panel of
Figure~\ref{fig:val-tot} shows the relative difference $\delta_{\rm
  NTS}$ at differential level.  Except for oscillations due to the
zero crossing of the distribution, the relative difference at the
level of the NNLO differential cross section, i.e.
\begin{equation}
\label{eq:val_nts}
  \frac{\D\sigma^{(2)}_{e\mu}}{\D\sigma_2}  \, \delta_{\rm NTS}
  \sim \alpha^2 \, \delta_{\rm NTS} \,,
\end{equation}
is at most around $10^{-2} \alpha^2$.  The error bars in this figure
only include the error of the real-virtual contributions. We stress
that this error is at least ten times smaller than the Monte Carlo
integration error of the physical NNLO correction, not shown in the
plot. Hence, we can safely argue that the NTS expansion is an
extremely good approximation.

As a consequence, the NTS expansion accurately represents the full
real-virtual contribution, as much as the full computation with {\sc
  OpenLoops} in quadruple precision. However, it allows for a much
faster integration.  The data in the plot, using the same amount of
statistics and the same setup for both cases, required seven days with
NTS stabilisation, three months without it.

%% file: 05_outlook.tex
\section{Conclusions and outlook}
\label{sec:conclusion}

We have presented the first calculation of the complete set of NNLO
QED corrections to muon-electron scattering. The results presented
here constitute the first complete and fully differential NNLO
calculation of a $2\to 2$ process with two different non-vanishing
masses on the external lines. They were obtained by combining
state-of-the-art analytic calculations of two-loop amplitudes and
their numerical evaluations with advanced QFT-inspired methods to
ensure an efficient and numerically stable phase-space integration,
even in the presence of a large hierarchy of scales. Through crossing,
they also open the door to study muon-pair and tau-pair production at
electron-positron colliders to unprecedented precision.  Furthermore,
they can be adapted to contribute to an improved description of
lepton-proton scattering.

Our results include leptonic, non-perturbative hadronic, and photonic
corrections. The fer\-mion\-ic matrix elements were calculated with
the semi-numerical hyperspherical method, while a purely analytic
approach was followed for the photonic parts. The phase-space
integration is performed numerically in the \mcmule{}
framework~\cite{Banerjee:2020rww}, where the soft singularities are
subtracted using the FKS$^\ell$ scheme~\cite{Engel:2019nfw}. This
allows for a fully-differential calculation of any IR-safe observable.

Electronic and muonic corrections have been computed exactly based on
the analytic result for the heavy-lepton and heavy-quark form
factors~\cite{Bonciani:2003ai,Bernreuther:2004ih}. The full mass
dependence of the genuine four-point two-loop topologies, on the other
hand, is currently not known. Adapting the recent computation of the
two-loop matrix element for
$e^+\,e^-\to\mu^+\mu^-$~\cite{Bonciani:2021okt} for massless
electrons, we have computed the corresponding matrix element for
muon-electron scattering. Based on this result, it was possible to
reconstruct the leading electron-mass effects using
massification~\cite{Engel:2018fsb}. This correctly captures both
logarithmically enhanced as well as constant terms and results within
an error of the order of $(\alpha/\pi)^2 m^2/S\log(m^2/S)$. It can
therefore be expected that the impact of the missing terms lies well
below MUonE's target precision of 10 ppm. We have substantiated this
claim by comparing the massive computation for the electron line
corrections with the massified result. The difference is smaller than
the Monte Carlo error, which is negligible for practical purposes.

The second bottleneck in the calculation is given by the numerical
(in)stability of the radiative matrix element for soft and collinear
photon emission. A stable and fast implementation of the delicate
real-virtual matrix element was possible by combining {\sc
  OpenLoops}~\cite{Buccioni:2017yxi, Buccioni:2019sur} with NTS
stabilisation~\cite{Banerjee:2021mty, Engel:2021ccn}. We have
cross-checked this approach by running {\sc OpenLoops} in
quadruple-precision mode and found perfect agreement of the two
results within the Monte Carlo error. Furthermore, the real-virtual
and double-real contributions have been validated in a dedicated
comparison of the radiative process with the {\sc Mesmer}
collaboration.

The calculation presented here represents a major milestone towards
the ambitious 10 ppm precision goal of the MUonE experiment. The
corresponding results unveil rather large NNLO corrections of up to
$\sim 10^{-3}$, depending on the angle of the scattered electron. They
are thus absolutely essential for the MUonE program. The enhancement
can be traced back to the presence of soft logarithms. An inclusion of
these effects beyond NNLO is therefore unavoidable to reach the
required precision. The leading logarithms can be resummed with a
parton shower and a corresponding effort is ongoing to extend the
\mcmule{} framework accordingly.

Even with a NNLO-matched parton shower it will be essential to
reliably assess the impact of missing higher-order contributions. An
analytic approach to resum next-to-leading logarithmic effects could
therefore be helpful in this regard. It is, however, unlikely that
this can be done for realistic MUonE observables. An alternative
approach is given by the fixed-order calculation of the electron-line
corrections at N$^3$LO. With the all-order subtraction scheme
FKS$^\ell$ and the recent calculation of the heavy-quark form factor
at three loop~\cite{Fael:2022miw, Fael:2022rgm} such an endeavour
indeed seems feasible. The main missing piece is the
real-virtual-virtual matrix element which is only known for massless
electrons~\cite{Garland:2001tf,Garland:2002ak}.  As an alternative
route to an analytic calculation of this matrix element, one could
follow a numerical approach to calculate the loop integrals using
e.g.~DiffExp~\cite{Hidding:2020ytt}, AMFlow~\cite{Liu:2022chg}, or
SecDec~\cite{Borowka:2015mxa}.  Another option is to extend
massification to radiative processes. A corresponding collaborative
effort has been launched recently~\cite{Durham:n3lo}.

Last but not least, a rigorous assessment of the reliability of the
massified approximation is clearly desirable. A computation with full
mass dependence is therefore envisaged. Analytic results for a subset
of the planar master integrals have recently become
available~\cite{Heller:2021gun}. A numerical approach seems, however,
more promising in this case. In fact, similar techniques as for the
electronic real-virtual-virtual amplitude might also be applicable
here. The N$^3$LO endeavour will therefore not only control missing
higher-order effects but might also facilitate the fully-massive NNLO
calculation. A prediction of the SM background for MUonE with the
ambitious target precision of 10 ppm thus seems within reach in the
near-term future.

%% file: mu-e-nnlo.bbl
\providecommand{\href}[2]{#2}\begingroup\raggedright\begin{thebibliography}{100}

\bibitem{Abbiendi:2016xup}
G.~Abbiendi et~al., \emph{{Measuring the leading hadronic contribution to the
  muon g-2 via $\mu e$ scattering}},
  \href{https://doi.org/10.1140/epjc/s10052-017-4633-z}{\emph{Eur. Phys. J. C}
  {\bfseries 77} (2017) 139}
  [\href{https://arxiv.org/abs/1609.08987}{{\ttfamily 1609.08987}}].

\bibitem{Spedicato:2022qtw}
{\scshape MUonE} collaboration, E.~Spedicato, \emph{{Status of the MUonE
  experiment}}, \href{https://doi.org/10.22323/1.398.0642}{\emph{PoS}
  {\bfseries EPS-HEP2021} (2022) 642}.

\bibitem{Abbiendi:2022oks}
G.~Abbiendi, \emph{{Status of the MUonE experiment}},
  \href{https://doi.org/10.1088/1402-4896/ac6297}{\emph{Phys. Scripta}
  {\bfseries 97} (2022) 054007}
  [\href{https://arxiv.org/abs/2201.13177}{{\ttfamily 2201.13177}}].

\bibitem{CarloniCalame:2015obs}
C.~M. Carloni~Calame, M.~Passera, L.~Trentadue and G.~Venanzoni, \emph{{A new
  approach to evaluate the leading hadronic corrections to the muon $g$-2}},
  \href{https://doi.org/10.1016/j.physletb.2015.05.020}{\emph{Phys. Lett. B}
  {\bfseries 746} (2015) 325}
  [\href{https://arxiv.org/abs/1504.02228}{{\ttfamily 1504.02228}}].

\bibitem{Muong-2:2006rrc}
{\scshape Muon g-2} collaboration, G.~W. Bennett et~al., \emph{{Final Report of
  the Muon E821 Anomalous Magnetic Moment Measurement at BNL}},
  \href{https://doi.org/10.1103/PhysRevD.73.072003}{\emph{Phys. Rev. D}
  {\bfseries 73} (2006) 072003}
  [\href{https://arxiv.org/abs/hep-ex/0602035}{{\ttfamily hep-ex/0602035}}].

\bibitem{Muong-2:2021ojo}
{\scshape Muon g-2} collaboration, B.~Abi et~al., \emph{{Measurement of the
  Positive Muon Anomalous Magnetic Moment to 0.46 ppm}},
  \href{https://doi.org/10.1103/PhysRevLett.126.141801}{\emph{Phys. Rev. Lett.}
  {\bfseries 126} (2021) 141801}
  [\href{https://arxiv.org/abs/2104.03281}{{\ttfamily 2104.03281}}].

\bibitem{Aoyama:2020ynm}
T.~Aoyama et~al., \emph{{The anomalous magnetic moment of the muon in the
  Standard Model}},
  \href{https://doi.org/10.1016/j.physrep.2020.07.006}{\emph{Phys. Rept.}
  {\bfseries 887} (2020) 1} [\href{https://arxiv.org/abs/2006.04822}{{\ttfamily
  2006.04822}}].

\bibitem{Davier:2019can}
M.~Davier, A.~Hoecker, B.~Malaescu and Z.~Zhang, \emph{{A new evaluation of the
  hadronic vacuum polarisation contributions to the muon anomalous magnetic
  moment and to $\alpha(m_Z^2)$}},
  \href{https://doi.org/10.1140/epjc/s10052-020-7792-2}{\emph{Eur. Phys. J. C}
  {\bfseries 80} (2020) 241}
  [\href{https://arxiv.org/abs/1908.00921}{{\ttfamily 1908.00921}}].

\bibitem{Keshavarzi:2019abf}
A.~Keshavarzi, D.~Nomura and T.~Teubner, \emph{{$g-2$ of charged leptons,
  $\alpha (M^2_Z)$ , and the hyperfine splitting of muonium}},
  \href{https://doi.org/10.1103/PhysRevD.101.014029}{\emph{Phys. Rev. D}
  {\bfseries 101} (2020) 014029}
  [\href{https://arxiv.org/abs/1911.00367}{{\ttfamily 1911.00367}}].

\bibitem{Borsanyi:2020mff}
S.~Borsanyi et~al., \emph{{Leading hadronic contribution to the muon magnetic
  moment from lattice QCD}},
  \href{https://doi.org/10.1038/s41586-021-03418-1}{\emph{Nature} {\bfseries
  593} (2021) 51} [\href{https://arxiv.org/abs/2002.12347}{{\ttfamily
  2002.12347}}].

\bibitem{Balzani:2021del}
E.~Balzani, S.~Laporta and M.~Passera, \emph{{Hadronic vacuum polarization
  contributions to the muon g-2 in the space-like region}},
  \href{https://doi.org/10.1016/j.physletb.2022.137462}{\emph{Phys. Lett. B}
  {\bfseries 834} (2022) 137462}
  [\href{https://arxiv.org/abs/2112.05704}{{\ttfamily 2112.05704}}].

\bibitem{Fael:2019nsf}
M.~Fael and M.~Passera, \emph{{Muon-Electron Scattering at
  Next-To-Next-To-Leading Order: The Hadronic Corrections}},
  \href{https://doi.org/10.1103/PhysRevLett.122.192001}{\emph{Phys. Rev. Lett.}
  {\bfseries 122} (2019) 192001}
  [\href{https://arxiv.org/abs/1901.03106}{{\ttfamily 1901.03106}}].

\bibitem{Fael:2018dmz}
M.~Fael, \emph{{Hadronic corrections to $\mu$-$e$ scattering at NNLO with
  space-like data}}, \href{https://doi.org/10.1007/JHEP02(2019)027}{\emph{JHEP}
  {\bfseries 02} (2019) 027}
  [\href{https://arxiv.org/abs/1808.08233}{{\ttfamily 1808.08233}}].

\bibitem{Masiero:2020vxk}
A.~Masiero, P.~Paradisi and M.~Passera, \emph{{New physics at the MUonE
  experiment at CERN}},
  \href{https://doi.org/10.1103/PhysRevD.102.075013}{\emph{Phys. Rev. D}
  {\bfseries 102} (2020) 075013}
  [\href{https://arxiv.org/abs/2002.05418}{{\ttfamily 2002.05418}}].

\bibitem{Dev:2020drf}
P.~S.~B. Dev, W.~Rodejohann, X.-J. Xu and Y.~Zhang, \emph{{MUonE sensitivity to
  new physics explanations of the muon anomalous magnetic moment}},
  \href{https://doi.org/10.1007/JHEP05(2020)053}{\emph{JHEP} {\bfseries 05}
  (2020) 053} [\href{https://arxiv.org/abs/2002.04822}{{\ttfamily
  2002.04822}}].

\bibitem{Schubert:2019nwm}
U.~Schubert and C.~Williams, \emph{{Interplay between SM precision, BSM
  physics, and the measurements of $\alpha_{\textrm{had}}$ in $\mu$-$e$
  scattering}}, \href{https://doi.org/10.1103/PhysRevD.100.035030}{\emph{Phys.
  Rev. D} {\bfseries 100} (2019) 035030}
  [\href{https://arxiv.org/abs/1907.01574}{{\ttfamily 1907.01574}}].

\bibitem{GrillidiCortona:2022kbq}
G.~Grilli~di Cortona and E.~Nardi, \emph{{Probing light mediators at the MUonE
  experiment}}, \href{https://doi.org/10.1103/PhysRevD.105.L111701}{\emph{Phys.
  Rev. D} {\bfseries 105} (2022) L111701}
  [\href{https://arxiv.org/abs/2204.04227}{{\ttfamily 2204.04227}}].

\bibitem{Galon:2022xcl}
I.~Galon, D.~Shih and I.~R. Wang, \emph{{Dark Photons and Displaced Vertices at
  the MUonE Experiment}},  \href{https://arxiv.org/abs/2202.08843}{{\ttfamily
  2202.08843}}.

\bibitem{Asai:2021wzx}
K.~Asai, K.~Hamaguchi, N.~Nagata, S.-Y. Tseng and J.~Wada, \emph{{Probing the
  $L_\mu$-$L_\tau$ gauge boson at the MUonE experiment}},
  \href{https://doi.org/10.1103/PhysRevD.106.L051702}{\emph{Phys. Rev. D}
  {\bfseries 106} (2022) L051702}
  [\href{https://arxiv.org/abs/2109.10093}{{\ttfamily 2109.10093}}].

\bibitem{Banerjee:2020tdt}
P.~Banerjee et~al., \emph{{Theory for muon-electron scattering @ 10 ppm: A
  report of the MUonE theory initiative}},
  \href{https://doi.org/10.1140/epjc/s10052-020-8138-9}{\emph{Eur. Phys. J. C}
  {\bfseries 80} (2020) 591}
  [\href{https://arxiv.org/abs/2004.13663}{{\ttfamily 2004.13663}}].

\bibitem{CarloniCalame:2020yoz}
C.~M. Carloni~Calame, M.~Chiesa, S.~M. Hasan, G.~Montagna, O.~Nicrosini and
  F.~Piccinini, \emph{{Towards muon-electron scattering at NNLO}},
  \href{https://doi.org/10.1007/JHEP11(2020)028}{\emph{JHEP} {\bfseries 11}
  (2020) 028} [\href{https://arxiv.org/abs/2007.01586}{{\ttfamily
  2007.01586}}].

\bibitem{Alacevich:2018vez}
M.~Alacevich, C.~M. Carloni~Calame, M.~Chiesa, G.~Montagna, O.~Nicrosini and
  F.~Piccinini, \emph{{Muon-electron scattering at NLO}},
  \href{https://doi.org/10.1007/JHEP02(2019)155}{\emph{JHEP} {\bfseries 02}
  (2019) 155} [\href{https://arxiv.org/abs/1811.06743}{{\ttfamily
  1811.06743}}].

\bibitem{Budassi:2021twh}
E.~Budassi, C.~M. Carloni~Calame, M.~Chiesa, C.~L. Del~Pio, S.~M. Hasan,
  G.~Montagna et~al., \emph{{NNLO virtual and real leptonic corrections to
  muon-electron scattering}},
  \href{https://doi.org/10.1007/JHEP11(2021)098}{\emph{JHEP} {\bfseries 11}
  (2021) 098} [\href{https://arxiv.org/abs/2109.14606}{{\ttfamily
  2109.14606}}].

\bibitem{Yennie:1961ad}
D.~R. Yennie, S.~C. Frautschi and H.~Suura, \emph{{The infrared divergence
  phenomena and high-energy processes}},
  \href{https://doi.org/10.1016/0003-4916(61)90151-8}{\emph{Annals Phys.}
  {\bfseries 13} (1961) 379}.

\bibitem{Budassi:2022kqs}
E.~Budassi, C.~M. Carloni~Calame, C.~L. Del~Pio and F.~Piccinini, \emph{{Single
  $\pi^0$ production in \ensuremath{\mu}e scattering at MUonE}},
  \href{https://doi.org/10.1016/j.physletb.2022.137138}{\emph{Phys. Lett. B}
  {\bfseries 829} (2022) 137138}
  [\href{https://arxiv.org/abs/2203.01639}{{\ttfamily 2203.01639}}].

\bibitem{Banerjee:2020rww}
P.~Banerjee, T.~Engel, A.~Signer and Y.~Ulrich, \emph{{QED at NNLO with
  McMule}}, \href{https://doi.org/10.21468/SciPostPhys.9.2.027}{\emph{SciPost
  Phys.} {\bfseries 9} (2020) 027}
  [\href{https://arxiv.org/abs/2007.01654}{{\ttfamily 2007.01654}}].

\bibitem{Engel:2019nfw}
T.~Engel, A.~Signer and Y.~Ulrich, \emph{{A subtraction scheme for massive
  QED}}, \href{https://doi.org/10.1007/JHEP01(2020)085}{\emph{JHEP} {\bfseries
  01} (2020) 085} [\href{https://arxiv.org/abs/1909.10244}{{\ttfamily
  1909.10244}}].

\bibitem{Frixione:1995ms}
S.~Frixione, Z.~Kunszt and A.~Signer, \emph{{Three jet cross-sections to
  next-to-leading order}},
  \href{https://doi.org/10.1016/0550-3213(96)00110-1}{\emph{Nucl. Phys. B}
  {\bfseries 467} (1996) 399}
  [\href{https://arxiv.org/abs/hep-ph/9512328}{{\ttfamily hep-ph/9512328}}].

\bibitem{Frederix:2009yq}
R.~Frederix, S.~Frixione, F.~Maltoni and T.~Stelzer, \emph{{Automation of
  next-to-leading order computations in QCD: The FKS subtraction}},
  \href{https://doi.org/10.1088/1126-6708/2009/10/003}{\emph{JHEP} {\bfseries
  10} (2009) 003} [\href{https://arxiv.org/abs/0908.4272}{{\ttfamily
  0908.4272}}].

\bibitem{Bonciani:2021okt}
R.~Bonciani et~al., \emph{{Two-Loop Four-Fermion Scattering Amplitude in QED}},
  \href{https://doi.org/10.1103/PhysRevLett.128.022002}{\emph{Phys. Rev. Lett.}
  {\bfseries 128} (2022) 022002}
  [\href{https://arxiv.org/abs/2106.13179}{{\ttfamily 2106.13179}}].

\bibitem{Mandal:2022vju}
M.~K. Mandal, P.~Mastrolia, J.~Ronca and W.~J. Bobadilla~Torres,
  \emph{{Two-loop scattering amplitude for heavy-quark pair production through
  light-quark annihilation in QCD}},
  \href{https://doi.org/10.1007/JHEP09(2022)129}{\emph{JHEP} {\bfseries 09}
  (2022) 129} [\href{https://arxiv.org/abs/2204.03466}{{\ttfamily
  2204.03466}}].

\bibitem{Mastrolia:2017pfy}
P.~Mastrolia, M.~Passera, A.~Primo and U.~Schubert, \emph{{Master integrals for
  the NNLO virtual corrections to $\mu e$ scattering in QED: the planar
  graphs}}, \href{https://doi.org/10.1007/JHEP11(2017)198}{\emph{JHEP}
  {\bfseries 11} (2017) 198}
  [\href{https://arxiv.org/abs/1709.07435}{{\ttfamily 1709.07435}}].

\bibitem{DiVita:2018nnh}
S.~Di~Vita, S.~Laporta, P.~Mastrolia, A.~Primo and U.~Schubert, \emph{{Master
  integrals for the NNLO virtual corrections to $\mu e$ scattering in QED: the
  non-planar graphs}},
  \href{https://doi.org/10.1007/JHEP09(2018)016}{\emph{JHEP} {\bfseries 09}
  (2018) 016} [\href{https://arxiv.org/abs/1806.08241}{{\ttfamily
  1806.08241}}].

\bibitem{Penin:2005eh}
A.~A. Penin, \emph{{Two-loop photonic corrections to massive Bhabha
  scattering}},
  \href{https://doi.org/10.1016/j.nuclphysb.2005.11.016}{\emph{Nucl. Phys. B}
  {\bfseries 734} (2006) 185}
  [\href{https://arxiv.org/abs/hep-ph/0508127}{{\ttfamily hep-ph/0508127}}].

\bibitem{Mitov:2006xs}
A.~Mitov and S.~Moch, \emph{{The Singular behavior of massive QCD amplitudes}},
  \href{https://doi.org/10.1088/1126-6708/2007/05/001}{\emph{JHEP} {\bfseries
  05} (2007) 001} [\href{https://arxiv.org/abs/hep-ph/0612149}{{\ttfamily
  hep-ph/0612149}}].

\bibitem{Becher:2007cu}
T.~Becher and K.~Melnikov, \emph{{Two-loop QED corrections to Bhabha
  scattering}},
  \href{https://doi.org/10.1088/1126-6708/2007/06/084}{\emph{JHEP} {\bfseries
  06} (2007) 084} [\href{https://arxiv.org/abs/0704.3582}{{\ttfamily
  0704.3582}}].

\bibitem{Engel:2018fsb}
T.~Engel, C.~Gnendiger, A.~Signer and Y.~Ulrich, \emph{{Small-mass effects in
  heavy-to-light form factors}},
  \href{https://doi.org/10.1007/JHEP02(2019)118}{\emph{JHEP} {\bfseries 02}
  (2019) 118} [\href{https://arxiv.org/abs/1811.06461}{{\ttfamily
  1811.06461}}].

\bibitem{Banerjee:2021mty}
P.~Banerjee, T.~Engel, N.~Schalch, A.~Signer and Y.~Ulrich, \emph{{Bhabha
  scattering at NNLO with next-to-soft stabilisation}},
  \href{https://doi.org/10.1016/j.physletb.2021.136547}{\emph{Phys. Lett. B}
  {\bfseries 820} (2021) 136547}
  [\href{https://arxiv.org/abs/2106.07469}{{\ttfamily 2106.07469}}].

\bibitem{Buccioni:2017yxi}
F.~Buccioni, S.~Pozzorini and M.~Zoller, \emph{{On-the-fly reduction of open
  loops}}, \href{https://doi.org/10.1140/epjc/s10052-018-5562-1}{\emph{Eur.
  Phys. J. C} {\bfseries 78} (2018) 70}
  [\href{https://arxiv.org/abs/1710.11452}{{\ttfamily 1710.11452}}].

\bibitem{Buccioni:2019sur}
F.~Buccioni, J.-N. Lang, J.~M. Lindert, P.~Maierh{\"{o}}fer, S.~Pozzorini,
  H.~Zhang et~al., \emph{{OpenLoops 2}},
  \href{https://doi.org/10.1140/epjc/s10052-019-7306-2}{\emph{Eur. Phys. J. C}
  {\bfseries 79} (2019) 866}
  [\href{https://arxiv.org/abs/1907.13071}{{\ttfamily 1907.13071}}].

\bibitem{Engel:2021ccn}
T.~Engel, A.~Signer and Y.~Ulrich, \emph{{Universal structure of radiative QED
  amplitudes at one loop}},
  \href{https://doi.org/10.1007/JHEP04(2022)097}{\emph{JHEP} {\bfseries 04}
  (2022) 097} [\href{https://arxiv.org/abs/2112.07570}{{\ttfamily
  2112.07570}}].

\bibitem{Low:1958sn}
F.~E. Low, \emph{{Bremsstrahlung of very low-energy quanta in elementary
  particle collisions}},
  \href{https://doi.org/10.1103/PhysRev.110.974}{\emph{Phys. Rev.} {\bfseries
  110} (1958) 974}.

\bibitem{Burnett:1967km}
T.~H. Burnett and N.~M. Kroll, \emph{{Extension of the low soft photon
  theorem}}, \href{https://doi.org/10.1103/PhysRevLett.20.86}{\emph{Phys. Rev.
  Lett.} {\bfseries 20} (1968) 86}.

\bibitem{Kollatzsch:2022bqa}
S.~Kollatzsch and Y.~Ulrich, \emph{{Lepton pair production at NNLO in QED with
  EW effects}},  \href{https://arxiv.org/abs/2210.17172}{{\ttfamily
  2210.17172}}.

\bibitem{Nikishov:1961}
A.~Nikishov, \emph{{Radiative corrections to the scattering of $\mu$ mesons on
  electrons}}, {\emph{Sov.Phys. JETP} {\bfseries 12} (1961) 529}.

\bibitem{Eriksson:1961}
K.~Eriksson, \emph{{Radiative corrections to muon-electron scattering}},
  {\emph{Nuovo Cim.} {\bfseries 19} (1961) 1029}.

\bibitem{Eriksson:1963}
K.~Eriksson, B.~Larsson and R.~G.A., \emph{{Radiative corrections to
  muon-electron scattering}}, {\emph{Nuovo Cim.} {\bfseries 30} (1963) 1434}.

\bibitem{VanNieuwenhuizen:1971yn}
P.~Van~Nieuwenhuizen, \emph{{Muon-electron scattering cross-section to order
  $\alpha^3$}}, \href{https://doi.org/10.1016/0550-3213(71)90009-5}{\emph{Nucl.
  Phys. B} {\bfseries 28} (1971) 429}.

\bibitem{Kukhto:1987uj}
T.~V. Kukhto, N.~M. Shumeiko and S.~I. Timoshin, \emph{{Radiative Corrections
  in Polarized Electron Muon Elastic Scattering}},
  \href{https://doi.org/10.1088/0305-4616/13/6/005}{\emph{J. Phys. G}
  {\bfseries 13} (1987) 725}.

\bibitem{Bardin:1997nc}
D.~Y. Bardin and L.~Kalinovskaya, \emph{{QED corrections for polarized elastic
  $\mu e$ scattering}},  \href{https://arxiv.org/abs/hep-ph/9712310}{{\ttfamily
  hep-ph/9712310}}.

\bibitem{Kaiser:2010zz}
N.~Kaiser, \emph{{Radiative corrections to lepton-lepton scattering
  revisited}}, \href{https://doi.org/10.1088/0954-3899/37/11/115005}{\emph{J.
  Phys. G} {\bfseries 37} (2010) 115005}.

\bibitem{Gnendiger:2017pys}
C.~Gnendiger et~al., \emph{{To ${d}$, or not to ${d}$: recent developments and
  comparisons of regularization schemes}},
  \href{https://doi.org/10.1140/epjc/s10052-017-5023-2}{\emph{Eur. Phys. J. C}
  {\bfseries 77} (2017) 471}
  [\href{https://arxiv.org/abs/1705.01827}{{\ttfamily 1705.01827}}].

\bibitem{Mastrolia:2003yz}
P.~Mastrolia and E.~Remiddi, \emph{{Two loop form-factors in QED}},
  \href{https://doi.org/10.1016/S0550-3213(03)00405-X}{\emph{Nucl. Phys. B}
  {\bfseries 664} (2003) 341}
  [\href{https://arxiv.org/abs/hep-ph/0302162}{{\ttfamily hep-ph/0302162}}].

\bibitem{Bonciani:2003ai}
R.~Bonciani, P.~Mastrolia and E.~Remiddi, \emph{{QED vertex form-factors at two
  loops}}, \href{https://doi.org/10.1016/j.nuclphysb.2003.10.031}{\emph{Nucl.
  Phys. B} {\bfseries 676} (2004) 399}
  [\href{https://arxiv.org/abs/hep-ph/0307295}{{\ttfamily hep-ph/0307295}}].

\bibitem{Bernreuther:2004ih}
W.~Bernreuther, R.~Bonciani, T.~Gehrmann, R.~Heinesch, T.~Leineweber,
  P.~Mastrolia et~al., \emph{{Two-loop QCD corrections to the heavy quark
  form-factors: The Vector contributions}},
  \href{https://doi.org/10.1016/j.nuclphysb.2004.10.059}{\emph{Nucl. Phys. B}
  {\bfseries 706} (2005) 245}
  [\href{https://arxiv.org/abs/hep-ph/0406046}{{\ttfamily hep-ph/0406046}}].

\bibitem{Banerjee:2021qvi}
P.~Banerjee, T.~Engel, N.~Schalch, A.~Signer and Y.~Ulrich, \emph{{M{\o}ller
  scattering at NNLO}},
  \href{https://doi.org/10.1103/PhysRevD.105.L031904}{\emph{Phys. Rev. D}
  {\bfseries 105} (2022) L031904}
  [\href{https://arxiv.org/abs/2107.12311}{{\ttfamily 2107.12311}}].

\bibitem{Engel:2022kde}
T.~Engel, \emph{{Muon-Electron Scattering at NNLO}},  {PhD Thesis},
  {Universit\"at Z\"urich}, 9, 2022.

\bibitem{Denner:2016kdg}
A.~Denner, S.~Dittmaier and L.~Hofer, \emph{{Collier: a fortran-based Complex
  One-Loop LIbrary in Extended Regularizations}},
  \href{https://doi.org/10.1016/j.cpc.2016.10.013}{\emph{Comput. Phys. Commun.}
  {\bfseries 212} (2017) 220}
  [\href{https://arxiv.org/abs/1604.06792}{{\ttfamily 1604.06792}}].

\bibitem{vanHameren:2010cp}
A.~van Hameren, \emph{{OneLOop: For the evaluation of one-loop scalar
  functions}}, \href{https://doi.org/10.1016/j.cpc.2011.06.011}{\emph{Comput.
  Phys. Commun.} {\bfseries 182} (2011) 2427}
  [\href{https://arxiv.org/abs/1007.4716}{{\ttfamily 1007.4716}}].

\bibitem{Levine:1974xh}
M.~J. Levine and R.~Roskies, \emph{{Hyperspherical approach to quantum
  electrodynamics - sixth-order magnetic moment}},
  \href{https://doi.org/10.1103/PhysRevD.9.421}{\emph{Phys. Rev. D} {\bfseries
  9} (1974) 421}.

\bibitem{Levine:1975jz}
M.~J. Levine, R.~C. Perisho and R.~Roskies, \emph{{Analytic Contributions to
  the G Factor of the electron}},
  \href{https://doi.org/10.1103/PhysRevD.13.997}{\emph{Phys. Rev. D} {\bfseries
  13} (1976) 997}.

\bibitem{Laporta:1994mb}
S.~Laporta, \emph{{Hyperspherical integration and the triple cross vertex
  graphs}}, \href{https://doi.org/10.1007/BF02780705}{\emph{Nuovo Cim. A}
  {\bfseries 107} (1994) 1729}
  [\href{https://arxiv.org/abs/hep-ph/9404203}{{\ttfamily hep-ph/9404203}}].

\bibitem{Djouadi:1993ss}
A.~Djouadi and P.~Gambino, \emph{{Electroweak gauge bosons selfenergies:
  Complete QCD corrections}},
  \href{https://doi.org/10.1103/PhysRevD.49.3499}{\emph{Phys. Rev. D}
  {\bfseries 49} (1994) 3499}
  [\href{https://arxiv.org/abs/hep-ph/9309298}{{\ttfamily hep-ph/9309298}}].

\bibitem{alphaqed}
F.~Jegerlehner. \url{http://www-com.physik.hu-berlin.de/~fjeger/software.html}.

\bibitem{Hahn:2000kx}
T.~Hahn, \emph{{Generating Feynman diagrams and amplitudes with FeynArts 3}},
  \href{https://doi.org/10.1016/S0010-4655(01)00290-9}{\emph{Comput. Phys.
  Commun.} {\bfseries 140} (2001) 418}
  [\href{https://arxiv.org/abs/hep-ph/0012260}{{\ttfamily hep-ph/0012260}}].

\bibitem{Shtabovenko:2016sxi}
V.~Shtabovenko, R.~Mertig and F.~Orellana, \emph{{New Developments in FeynCalc
  9.0}}, \href{https://doi.org/10.1016/j.cpc.2016.06.008}{\emph{Comput. Phys.
  Commun.} {\bfseries 207} (2016) 432}
  [\href{https://arxiv.org/abs/1601.01167}{{\ttfamily 1601.01167}}].

\bibitem{Mastrolia:2019aid}
P.~Mastrolia, T.~Peraro, A.~Primo, J.~Ronca and W.~J. Torres~Bobadilla,
  \emph{{AIDA, Adaptive Integrand Decomposition Algorithm}}.

\bibitem{Mastrolia:2016dhn}
P.~Mastrolia, T.~Peraro and A.~Primo, \emph{{Adaptive Integrand Decomposition
  in parallel and orthogonal space}},
  \href{https://doi.org/10.1007/JHEP08(2016)164}{\emph{JHEP} {\bfseries 08}
  (2016) 164} [\href{https://arxiv.org/abs/1605.03157}{{\ttfamily
  1605.03157}}].

\bibitem{Mastrolia:2016xxx}
P.~Mastrolia, A.~Primo and W.~J. Torres~Bobadilla, \emph{{Multi-gluon
  Scattering Amplitudes at One-Loop and Color-Kinematics Duality}}, {\emph{In
  preparation} (2016)}.

\bibitem{vonManteuffel:2012np}
A.~von Manteuffel and C.~Studerus, \emph{{Reduze 2 - Distributed Feynman
  Integral Reduction}},  \href{https://arxiv.org/abs/1201.4330}{{\ttfamily
  1201.4330}}.

\bibitem{Barucchi:1973zm}
G.~Barucchi and G.~Ponzano, \emph{{Differential equations for one-loop
  generalized feynman integrals}},
  \href{https://doi.org/10.1063/1.1666327}{\emph{J. Math. Phys.} {\bfseries 14}
  (1973) 396}.

\bibitem{Kotikov:1990kg}
A.~V. Kotikov, \emph{{Differential equations method: New technique for massive
  Feynman diagrams calculation}},
  \href{https://doi.org/10.1016/0370-2693(91)90413-K}{\emph{Phys. Lett. B}
  {\bfseries 254} (1991) 158}.

\bibitem{Remiddi:1997ny}
E.~Remiddi, \emph{{Differential equations for Feynman graph amplitudes}},
  \href{https://doi.org/10.1007/BF03185566}{\emph{Nuovo Cim. A} {\bfseries 110}
  (1997) 1435} [\href{https://arxiv.org/abs/hep-th/9711188}{{\ttfamily
  hep-th/9711188}}].

\bibitem{Gehrmann:1999as}
T.~Gehrmann and E.~Remiddi, \emph{{Differential equations for two loop four
  point functions}},
  \href{https://doi.org/10.1016/S0550-3213(00)00223-6}{\emph{Nucl. Phys. B}
  {\bfseries 580} (2000) 485}
  [\href{https://arxiv.org/abs/hep-ph/9912329}{{\ttfamily hep-ph/9912329}}].

\bibitem{Henn:2013pwa}
J.~M. Henn, \emph{{Multiloop integrals in dimensional regularization made
  simple}}, \href{https://doi.org/10.1103/PhysRevLett.110.251601}{\emph{Phys.
  Rev. Lett.} {\bfseries 110} (2013) 251601}
  [\href{https://arxiv.org/abs/1304.1806}{{\ttfamily 1304.1806}}].

\bibitem{Argeri:2014qva}
M.~Argeri, S.~Di~Vita, P.~Mastrolia, E.~Mirabella, J.~Schlenk, U.~Schubert
  et~al., \emph{{Magnus and Dyson Series for Master Integrals}},
  \href{https://doi.org/10.1007/JHEP03(2014)082}{\emph{JHEP} {\bfseries 03}
  (2014) 082} [\href{https://arxiv.org/abs/1401.2979}{{\ttfamily 1401.2979}}].

\bibitem{DiVita:2014pza}
S.~Di~Vita, P.~Mastrolia, U.~Schubert and V.~Yundin, \emph{{Three-loop master
  integrals for ladder-box diagrams with one massive leg}},
  \href{https://doi.org/10.1007/JHEP09(2014)148}{\emph{JHEP} {\bfseries 09}
  (2014) 148} [\href{https://arxiv.org/abs/1408.3107}{{\ttfamily 1408.3107}}].

\bibitem{Goncharov:1998kja}
A.~B. Goncharov, \emph{{Multiple polylogarithms, cyclotomy and modular
  complexes}}, \href{https://doi.org/10.4310/MRL.1998.v5.n4.a7}{\emph{Math.
  Res. Lett.} {\bfseries 5} (1998) 497}
  [\href{https://arxiv.org/abs/1105.2076}{{\ttfamily 1105.2076}}].

\bibitem{Gehrmann:2001jv}
T.~Gehrmann and E.~Remiddi, \emph{{Numerical evaluation of two-dimensional
  harmonic polylogarithms}},
  \href{https://doi.org/10.1016/S0010-4655(02)00139-X}{\emph{Comput. Phys.
  Commun.} {\bfseries 144} (2002) 200}
  [\href{https://arxiv.org/abs/hep-ph/0111255}{{\ttfamily hep-ph/0111255}}].

\bibitem{Czakon:2008zk}
M.~Czakon, \emph{{Tops from Light Quarks: Full Mass Dependence at Two-Loops in
  QCD}}, \href{https://doi.org/10.1016/j.physletb.2008.05.028}{\emph{Phys.
  Lett. B} {\bfseries 664} (2008) 307}
  [\href{https://arxiv.org/abs/0803.1400}{{\ttfamily 0803.1400}}].

\bibitem{Bonciani:2008az}
R.~Bonciani, A.~Ferroglia, T.~Gehrmann, D.~Maitre and C.~Studerus,
  \emph{{Two-Loop Fermionic Corrections to Heavy-Quark Pair Production: The
  Quark-Antiquark Channel}},
  \href{https://doi.org/10.1088/1126-6708/2008/07/129}{\emph{JHEP} {\bfseries
  07} (2008) 129} [\href{https://arxiv.org/abs/0806.2301}{{\ttfamily
  0806.2301}}].

\bibitem{Bonciani:2009nb}
R.~Bonciani, A.~Ferroglia, T.~Gehrmann and C.~Studerus, \emph{{Two-Loop Planar
  Corrections to Heavy-Quark Pair Production in the Quark-Antiquark Channel}},
  \href{https://doi.org/10.1088/1126-6708/2009/08/067}{\emph{JHEP} {\bfseries
  08} (2009) 067} [\href{https://arxiv.org/abs/0906.3671}{{\ttfamily
  0906.3671}}].

\bibitem{Nogueira:1991ex}
P.~Nogueira, \emph{{Automatic Feynman graph generation}},
  \href{https://doi.org/10.1006/jcph.1993.1074}{\emph{J. Comput. Phys.}
  {\bfseries 105} (1993) 279}.

\bibitem{Patel:2015tea}
H.~H. Patel, \emph{{Package-X: A Mathematica package for the analytic
  calculation of one-loop integrals}},
  \href{https://doi.org/10.1016/j.cpc.2015.08.017}{\emph{Comput. Phys. Commun.}
  {\bfseries 197} (2015) 276}
  [\href{https://arxiv.org/abs/1503.01469}{{\ttfamily 1503.01469}}].

\bibitem{Becher:2009kw}
T.~Becher and M.~Neubert, \emph{{Infrared singularities of QCD amplitudes with
  massive partons}},
  \href{https://doi.org/10.1103/PhysRevD.79.125004}{\emph{Phys. Rev. D}
  {\bfseries 79} (2009) 125004}
  [\href{https://arxiv.org/abs/0904.1021}{{\ttfamily 0904.1021}}].

\bibitem{Becher:2009qa}
T.~Becher and M.~Neubert, \emph{{On the Structure of Infrared Singularities of
  Gauge-Theory Amplitudes}},
  \href{https://doi.org/10.1088/1126-6708/2009/06/081}{\emph{JHEP} {\bfseries
  06} (2009) 081} [\href{https://arxiv.org/abs/0903.1126}{{\ttfamily
  0903.1126}}].

\bibitem{Hill:2016gdf}
R.~J. Hill, \emph{{Effective field theory for large logarithms in radiative
  corrections to electron proton scattering}},
  \href{https://doi.org/10.1103/PhysRevD.95.013001}{\emph{Phys. Rev. D}
  {\bfseries 95} (2017) 013001}
  [\href{https://arxiv.org/abs/1605.02613}{{\ttfamily 1605.02613}}].

\bibitem{Broggio:2015dga}
A.~Broggio, C.~Gnendiger, A.~Signer, D.~St\"ockinger and A.~Visconti,
  \emph{{SCET approach to regularization-scheme dependence of QCD amplitudes}},
  \href{https://doi.org/10.1007/JHEP01(2016)078}{\emph{JHEP} {\bfseries 01}
  (2016) 078} [\href{https://arxiv.org/abs/1506.05301}{{\ttfamily
  1506.05301}}].

\bibitem{Gnendiger:2016cpg}
C.~Gnendiger, A.~Signer and A.~Visconti, \emph{{Regularization-scheme
  dependence of QCD amplitudes in the massive case}},
  \href{https://doi.org/10.1007/JHEP10(2016)034}{\emph{JHEP} {\bfseries 10}
  (2016) 034} [\href{https://arxiv.org/abs/1607.08241}{{\ttfamily
  1607.08241}}].

\bibitem{Kilgore:2012tb}
W.~B. Kilgore, \emph{{The Four Dimensional Helicity Scheme Beyond One Loop}},
  \href{https://doi.org/10.1103/PhysRevD.86.014019}{\emph{Phys. Rev. D}
  {\bfseries 86} (2012) 014019}
  [\href{https://arxiv.org/abs/1205.4015}{{\ttfamily 1205.4015}}].

\bibitem{Gnendiger:2014nxa}
C.~Gnendiger, A.~Signer and D.~St\"ockinger, \emph{{The infrared structure of
  QCD amplitudes and $H \to gg$ in FDH and DRED}},
  \href{https://doi.org/10.1016/j.physletb.2014.05.003}{\emph{Phys. Lett. B}
  {\bfseries 733} (2014) 296}
  [\href{https://arxiv.org/abs/1404.2171}{{\ttfamily 1404.2171}}].

\bibitem{Ulrich:2020frs}
Y.~Ulrich, \emph{{McMule -- QED Corrections for Low-Energy Experiments}},  {PhD
  Thesis}, {Universit\"at Z\"urich}, 8, 2020.

\bibitem{Bauer:2000yr}
C.~W. Bauer, S.~Fleming, D.~Pirjol and I.~W. Stewart, \emph{{An Effective field
  theory for collinear and soft gluons: Heavy to light decays}},
  \href{https://doi.org/10.1103/PhysRevD.63.114020}{\emph{Phys. Rev. D}
  {\bfseries 63} (2001) 114020}
  [\href{https://arxiv.org/abs/hep-ph/0011336}{{\ttfamily hep-ph/0011336}}].

\bibitem{Bauer:2001yt}
C.~W. Bauer, D.~Pirjol and I.~W. Stewart, \emph{{Soft collinear factorization
  in effective field theory}},
  \href{https://doi.org/10.1103/PhysRevD.65.054022}{\emph{Phys. Rev. D}
  {\bfseries 65} (2002) 054022}
  [\href{https://arxiv.org/abs/hep-ph/0109045}{{\ttfamily hep-ph/0109045}}].

\bibitem{Beneke:2002ph}
M.~Beneke, A.~P. Chapovsky, M.~Diehl and T.~Feldmann, \emph{{Soft collinear
  effective theory and heavy to light currents beyond leading power}},
  \href{https://doi.org/10.1016/S0550-3213(02)00687-9}{\emph{Nucl. Phys. B}
  {\bfseries 643} (2002) 431}
  [\href{https://arxiv.org/abs/hep-ph/0206152}{{\ttfamily hep-ph/0206152}}].

\bibitem{Beneke:1997zp}
M.~Beneke and V.~A. Smirnov, \emph{{Asymptotic expansion of Feynman integrals
  near threshold}},
  \href{https://doi.org/10.1016/S0550-3213(98)00138-2}{\emph{Nucl. Phys. B}
  {\bfseries 522} (1998) 321}
  [\href{https://arxiv.org/abs/hep-ph/9711391}{{\ttfamily hep-ph/9711391}}].

\bibitem{mcmule}
\mcmule{} Team. \url{https://mule-tools.gitlab.io/}.

\bibitem{Naterop:2019xaf}
L.~Naterop, A.~Signer and Y.~Ulrich, \emph{{handyG \textemdash{}Rapid numerical
  evaluation of generalised polylogarithms in Fortran}},
  \href{https://doi.org/10.1016/j.cpc.2020.107165}{\emph{Comput. Phys. Commun.}
  {\bfseries 253} (2020) 107165}
  [\href{https://arxiv.org/abs/1909.01656}{{\ttfamily 1909.01656}}].

\bibitem{McMule:data}
\mcmule{} Team, ``\mcmule{} dataset.''
  \url{https://doi.org/10.5281/zenodo.6541686}.

\bibitem{Lannelongue:2021co2}
L.~Lannelongue, J.~Grealey and M.~Inouye, \emph{Green algorithms: Quantifying
  the carbon footprint of computation},
  \href{https://doi.org/https://doi.org/10.1002/advs.202100707}{\emph{Advanced
  Science} {\bfseries 8} (2021) 2100707}
  [\href{https://arxiv.org/abs/https://onlinelibrary.wiley.com/doi/pdf/10.1002/advs.202100707}{{\ttfamily
  https://onlinelibrary.wiley.com/doi/pdf/10.1002/advs.202100707}}].

\bibitem{Workman:2022ynf}
{\scshape Particle Data Group} collaboration, R.~L. Workman and Others,
  \emph{{Review of Particle Physics}},
  \href{https://doi.org/10.1093/ptep/ptac097}{\emph{PTEP} {\bfseries 2022}
  (2022) 083C01}.

\bibitem{Bern:2000ie}
Z.~Bern, L.~J. Dixon and A.~Ghinculov, \emph{{Two loop correction to Bhabha
  scattering}}, \href{https://doi.org/10.1103/PhysRevD.63.053007}{\emph{Phys.
  Rev. D} {\bfseries 63} (2001) 053007}
  [\href{https://arxiv.org/abs/hep-ph/0010075}{{\ttfamily hep-ph/0010075}}].

\bibitem{Fael:2022miw}
M.~Fael, F.~Lange, K.~Sch\"onwald and M.~Steinhauser, \emph{{Singlet and
  nonsinglet three-loop massive form factors}},
  \href{https://doi.org/10.1103/PhysRevD.106.034029}{\emph{Phys. Rev. D}
  {\bfseries 106} (2022) 034029}
  [\href{https://arxiv.org/abs/2207.00027}{{\ttfamily 2207.00027}}].

\bibitem{Fael:2022rgm}
M.~Fael, F.~Lange, K.~Sch\"onwald and M.~Steinhauser, \emph{{Massive Vector
  Form Factors to Three Loops}},
  \href{https://doi.org/10.1103/PhysRevLett.128.172003}{\emph{Phys. Rev. Lett.}
  {\bfseries 128} (2022) 172003}
  [\href{https://arxiv.org/abs/2202.05276}{{\ttfamily 2202.05276}}].

\bibitem{Garland:2001tf}
L.~W. Garland, T.~Gehrmann, E.~W.~N. Glover, A.~Koukoutsakis and E.~Remiddi,
  \emph{{The Two loop QCD matrix element for $e^+e^- \to 3~\text{Jets}$}},
  \href{https://doi.org/10.1016/S0550-3213(02)00057-3}{\emph{Nucl. Phys. B}
  {\bfseries 627} (2002) 107}
  [\href{https://arxiv.org/abs/hep-ph/0112081}{{\ttfamily hep-ph/0112081}}].

\bibitem{Garland:2002ak}
L.~W. Garland, T.~Gehrmann, E.~W.~N. Glover, A.~Koukoutsakis and E.~Remiddi,
  \emph{{Two loop QCD helicity amplitudes for $e^+e^- \to 3~\text{Jets}$}},
  \href{https://doi.org/10.1016/S0550-3213(02)00627-2}{\emph{Nucl. Phys. B}
  {\bfseries 642} (2002) 227}
  [\href{https://arxiv.org/abs/hep-ph/0206067}{{\ttfamily hep-ph/0206067}}].

\bibitem{Hidding:2020ytt}
M.~Hidding, \emph{{DiffExp, a Mathematica package for computing Feynman
  integrals in terms of one-dimensional series expansions}},
  \href{https://doi.org/10.1016/j.cpc.2021.108125}{\emph{Comput. Phys. Commun.}
  {\bfseries 269} (2021) 108125}
  [\href{https://arxiv.org/abs/2006.05510}{{\ttfamily 2006.05510}}].

\bibitem{Liu:2022chg}
X.~Liu and Y.-Q. Ma, \emph{{AMFlow: A Mathematica package for Feynman integrals
  computation via auxiliary mass flow}},
  \href{https://doi.org/10.1016/j.cpc.2022.108565}{\emph{Comput. Phys. Commun.}
  {\bfseries 283} (2023) 108565}
  [\href{https://arxiv.org/abs/2201.11669}{{\ttfamily 2201.11669}}].

\bibitem{Borowka:2015mxa}
S.~Borowka, G.~Heinrich, S.~P. Jones, M.~Kerner, J.~Schlenk and T.~Zirke,
  \emph{{SecDec-3.0: numerical evaluation of multi-scale integrals beyond one
  loop}}, \href{https://doi.org/10.1016/j.cpc.2015.05.022}{\emph{Comput. Phys.
  Commun.} {\bfseries 196} (2015) 470}
  [\href{https://arxiv.org/abs/1502.06595}{{\ttfamily 1502.06595}}].

\bibitem{Durham:n3lo}
Y.~Ulrich, ``{N$^3$LO kick-off workstop/thinkstart, Durham, 3--5 August,
  2022}.'' \url{https://conference.ippp.dur.ac.uk/event/1104/}.

\bibitem{Heller:2021gun}
M.~Heller, \emph{{Planar two-loop integrals for $\mu e$ scattering in QED with
  finite lepton masses}},  \href{https://arxiv.org/abs/2105.08046}{{\ttfamily
  2105.08046}}.

\end{thebibliography}\endgroup
